\documentclass[sigconf,nonacm,screen]{acmart}
\usepackage{cleveref}
\usepackage{graphicx}
\usepackage{booktabs,colortbl,multicol,multirow}
\usepackage[table]{xcolor}

\usepackage{fontawesome}
\usepackage[most]{tcolorbox}
\usepackage{tikz}
\usetikzlibrary{arrows.meta,positioning,shapes.geometric}

\newtcolorbox{takeawaybox}{
    enhanced,
    colback=black!4,
    frame hidden,
    boxrule=0pt,
    borderline west={3pt}{0pt}{blue!75!black},
    left=3mm,
    right=3mm,
    top=2mm,
    bottom=2mm,
    sharp corners
}

\newcommand{\system}[0]{\textsc{Pilot}}
\newcommand{\signals}[0]{\textsc{Surface}$_\text{A}$}
\newcommand{\statements}[0]{\textsc{Surface}$_\text{B}$}
\newcommand{\structure}[0]{\textsc{Surface}$_\text{C}$}
\newcommand{\surfaceA}[0]{\textsc{Surface}$_\text{A}$}
\newcommand{\surfaceB}[0]{\textsc{Surface}$_\text{B}$}
\newcommand{\surfaceC}[0]{\textsc{Surface}$_\text{C}$}

\AtBeginDocument{%
  }

\begin{document}
\sloppy

\title{\includegraphics[height=1.2em]{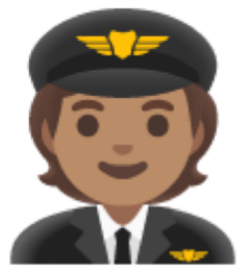} \system{}: Control Surfaces for Authoring Social Media Feeds}

\author{Agam Goyal}
\authornote{Both authors contributed equally}
\orcid{0009-0009-5989-2887}
\affiliation{%
  \institution{University of Illinois Urbana-Champaign}
  \city{Urbana}
  \state{Illinois}
  \country{USA}}
\email{agamg2@illinois.edu}

\author{Frederick Choi}
\authornotemark[1]
\orcid{0000-0002-8818-2456}
\affiliation{%
  \institution{University of Illinois Urbana-Champaign}
  \city{Urbana}
  \state{IL}
  \country{USA}
}
\email{fc20@illinois.edu}

\author{Eshwar Chandrasekharan}
\orcid{0000-0002-7473-1418}
\affiliation{%
  \institution{University of Illinois Urbana-Champaign}
  \city{Urbana}
  \state{Illinois}
  \country{USA}}
\email{eshwar@illinois.edu}


\definecolor{DarkBlue}{HTML}{00008B}
\definecolor{mscolor}{HTML}{01665e}
\definecolor{nmscolor}{HTML}{bf812d}
\definecolor{lgreen}{HTML}{ccece6}
\definecolor{dolive}{HTML}{308014}
\definecolor{purple}{HTML}{ae017e}
\definecolor{brickred}{HTML}{f03b20}

\definecolor{editCol}{HTML}{01665e}
\definecolor{maskCol}{HTML}{bf812d}

\newcommand{\edit}[1]{{\textcolor{editCol}{#1}}}
\newcommand{\mask}[1]{{\textcolor{maskCol}{#1}}}
\newcommand{\para}[1]{\noindent\textbf{\textit{#1}~}}

% Participant block quotes. \pquote{text}{C3}, \pquotetwo pairs two contrasting speakers.
\newcommand{\pquote}[2]{%
  \begin{quote}
    \emph{``#1''}~\textnormal{(#2)}
  \end{quote}}
\newcommand{\pquotetwo}[4]{%
  \begin{quote}
    \emph{``#1''}~\textnormal{(#2)}
    \par\smallskip\noindent
    \emph{``#3''}~\textnormal{(#4)}
  \end{quote}}
\newcommand{\takeaway}[1]{%
  \par\medskip\noindent\textbf{Takeaway.}~#1\par\medskip}

\renewcommand{\shortauthors}{Goyal and Choi et al.}

\begin{abstract}
Personalized social media feeds infer preferences from behavior, leaving people little direct control over what they see. Existing controls range from post-level reactions to rules and natural language, but little is known about how people use them together or how added expressiveness changes effort. We built \system{}, a Bluesky feed-authoring system that turns in-feed actions into explicit preferences, deterministic ranking, and inspectable outcomes. A study with seven participants informed an expanded implementation, which we organized into three nested control surfaces that ten participants compared within subjects. Participants assigned controls to distinct jobs: broad controls set direction, in-post actions refined results, and filters removed content. Richer surfaces did not necessarily feel more effortful, and participants' experiences depended more on whether they could verify and repair outcomes. Our findings show that usable feed control requires not simply more controls, but orchestration across mechanisms that support expression, inspection, repair, and episodic use.
\end{abstract}




\maketitle

\section{Introduction}

Social media feeds bring news, entertainment, and public discourse into a single stream drawn from more available content than any one person can inspect. Platforms therefore rely on personalized ranking to decide which posts appear and in what order, commonly learning from behavioral traces such as likes and dwell time~\cite{narayananUnderstandingSocialMedia2023,fengMappingDesignSpace2024}. Although personalization reduces this volume, it also gives ranking systems a consequential role in shaping what people encounter. Features such as likes, comments, and shares now serve both interpersonal purposes and algorithmic personalization, roles that recent work describes as entangled~\cite{10.1145/3772318.3791252}. How these actions alter the feed often remains invisible~\cite{eslamiAlwaysAssumedThat2015,eslamiFirstItThen2016}, leaving users to rely on folk theories and strategic interactions whose efficacy is unclear~\cite{devitoHowPeopleForm2018,karizatAlgorithmicFolkTheories2021,simpsonHowTameYour2022,fengMappingDesignSpace2024}.

Platforms and researchers have responded with explicit controls. These range from per-post feedback and Bluesky custom feeds~\cite{fengMappingDesignSpace2024,theblueskyteamAlgorithmicChoiceCustom2023,kleppmannBlueskyProtocolUsable2024} to rule authoring and natural language specification~\cite{wangEndUserAuthoring2025,10.1145/3772318.3791855,barua2026compass}. This direction connects interactive machine teaching, which treats users as teachers who iteratively correct a system \cite{ramosInteractiveMachineTeaching2020,fengMappingDesignSpace2024}, with work on value-aligned feeds that lets users state objectives beyond engagement~\cite{10.1145/3772318.3791281,goyal2026uncovering}. Both areas depend on usable ways to state and revise intent, yet prior studies show that higher levels of control can increase cognitive load~\cite{jinHowDifferentLevels2017}, feed settings can be difficult to find~\cite{hsuAwarenessNavigationUse2020}, and even natural language feed construction can demand more effort than users are accustomed to~\cite{10.1145/3772318.3791855}. What remains understudied is what happens when these mechanisms coexist: which one a person reaches for, what the added expressiveness costs, and how they recover when an interpretation goes wrong.

We address this problem by building \system{}, a working feed-authoring system on top of the Bluesky client and API. Its controls sit within a live feed so users can act on the content that prompted a reaction. Rather than inferring a latent profile from ordinary behavior, \system{} records explicit preferences and evaluates them through a deterministic ranker. Its shared architecture keeps actions, stored preferences, outcomes, and explanations traceable while allowing us to vary the available control surface. This makes it possible to study not only what people express, but also where effort enters, how they recognize and repair misalignment, and which surface they select while browsing real content.

We first evaluated an initial version of \system{} in which people authored feeds through direct controls available within the feed (Study 1). Seven participants chose sources, wrote keyword and author filters, set priority weights, and read the resulting feeds during open-ended sessions. This study addresses the following research questions:
\begin{itemize}
    \item[\textbf{RQ1a}] What do people attempt to express through direct feed controls, and which controls do they reach for?
    \item[\textbf{RQ1b}] Where do people's intentions exceed what structured controls can express?
    \item[\textbf{RQ1c}] How does feed authoring fit alongside ordinary browsing, and when do people engage the controls?
\end{itemize}

Participants treated filters as hard constraints and weights as soft preferences, while reactive exclusions often began from the control beside the post that prompted them. Their intentions also revealed gaps between literal tokens and the meanings they wanted to express. Some participants could describe a preference in a sentence but could not find a keyword that captured it, and some valued being able to put the controls away when they returned from authoring to reading. These findings, participant requests, and growing interest in language models as interfaces for feed curation motivated a wider set of capabilities. We used this evidence to add inferred topics that match meaning rather than tokens, a visible and editable preference store, natural language requests that compile into reviewable rules, per-post provenance, and reversible history. Language models propose metadata and candidate rules for users to accept, but do not score or rerank posts.

Study 1, however, could not tell us whether addressing these observed gaps with richer controls would improve users' ability to express, understand, and repair their preferences, what additional effort those capabilities would impose, or how people would combine them when several were available. We therefore use Study 2 to characterize these tradeoffs by instantiating three nested control surfaces of \system{} in order to compare different levels of control (Study 2). \surfaceA{} offers post-level preference buttons, \surfaceB{} adds keyword and author filters, natural language requests, and a visible preference store, and \surfaceC{} adds semantic inference, structured editing, weights, in-post topic chips, provenance, and history. We kept the ranking procedure and prepared sources fixed so that the comparison would isolate the surfaces rather than the ranker, and we used the nesting to study what each surface changes rather than to position the richest one as the intended endpoint. This study addresses the following research questions:
\begin{itemize}
    \item[\textbf{RQ2a}] How does what people express change as the control surface has more layers?
    \item[\textbf{RQ2b}] What does each additional layer cost in effort, predictability, and willingness to curate?
    \item[\textbf{RQ2c}] How do people detect and repair misalignment, and how does this differ across control surfaces?
    \item[\textbf{RQ2d}] When all layers are available, which do people use for which purposes?
\end{itemize}

Through our work, we contribute (i) an account of what people express through direct feed controls and where their intentions exceed literal rules, (ii) \system{}\footnote{Code to enable use of \system{} on Bluesky, and a system demonstration will be made available publicly upon acceptance.}, a working feed-authoring system that unifies in-feed actions, a shared preference store, and inspectable outcomes, and (iii) a comparative characterization of how people express, verify, and repair preferences across three nested control surfaces, showing how these surfaces support complementary roles rather than forming a simple progression toward greater control. Together, these contributions show that richer surfaces can sometimes feel less effortful and identify needs that none of the surfaces yet support.

\subsection{Summary of Findings}

Control surfaces differed less in how much control they offered than in which part of the judgment the user had to supply, which shaped what they could express and how they could repair mistakes. Per-post feedback conveyed a judgment without its reason, and seven participants encountered an incorrect inference about why they reacted. Richer surfaces changed which reasons participants could name rather than merely how much they could say. Felt effort also did not rise with the number of controls, as several participants found a less capable surface more demanding and linked trust more closely to visible follow-through and repair than to expressiveness. When all layers were available, participants used a coarse-to-fine workflow, reacting while reading and opening the preference store later to inspect and repair.

\subsection{Summary of Implications}

Our findings indicate that the design problem for feed authoring and curation systems is not just which features to ship, but how to orchestrate several of them over time. Showing every control at once can transfer work onto the user, so systems should offer the next useful action rather than the full suite. Cheap reactions also create preference debt, the later work of cleaning stored rules that no longer match what the user wants once their effects combine. Preference stores therefore need maintenance lifecycle support beyond inspectability. Controls should also support episodic curation by remaining accessible without being permanently visible. Finally, as feed curation moves toward being agentic, having a persistent memory would be crucial, and those memories should be inspectable, editable, and faithful to ranking. 

\section{Related Work}

Our work sits at the intersection of research on user agency in personalized feeds, end-user mechanisms for explicitly authoring personalization, and interactive systems for inspecting and correcting what a system has learned.

\subsection{Agency, Intent, and Personalized Feeds}

Personalized ranking helps people navigate volumes of social media content that they could not inspect directly, but it also determines what receives attention and what remains unseen~\cite{narayananUnderstandingSocialMedia2023,strayBuildingHumanValues2024}. Most deployed systems learn preferences primarily from behavioral signals such as viewing, liking, sharing, and commenting. These signals are useful for prediction but need not correspond to what a person deliberately wants from their feed. Moreover, their effect on subsequent recommendations is often difficult to observe. Some users are unaware that their feeds are algorithmically curated~\cite{eslamiFirstItThen2016,granBeNotBe2021}, while others develop folk theories and strategically alter their behavior to influence the algorithm~\cite{devitoHowPeopleForm2018,karizatAlgorithmicFolkTheories2021,simpsonHowTameYour2022}. Such strategies can include avoiding content a person would otherwise engage with for fear of changing future recommendations~\cite{simpsonHowTameYour2022}.

This motivates a shift from inferring preferences exclusively from behavior toward giving people ways to state what they want. Prior work has found that user control can support satisfaction, understanding, and perceived agency~\cite{burrellWhenUsersControl2019,harambamDesigningBetterTaking2019}. Work on recommender systems has similarly distinguished between indirect behavioral feedback and mechanisms that let users intervene more directly in personalization. For example, \citet{harperPuttingUsersControl2015} gave users explicit controls for steering recommendations and found substantial variation in how people chose to adjust the same underlying recommender. Related work on value-aligned recommendation similarly argues that feed objectives need not be limited to engagement and can instead reflect higher-level goals and values~\cite{strayBuildingHumanValues2024,10.1145/3772318.3791281,goyal2026uncovering}, and notions of desirability in online settings~\cite{goyal2026language,lambert2025does,lambert2025mind}. These approaches, however, still depend on interfaces through which people can express and revise those goals. \textit{Our work focuses on this interaction layer: how different mechanisms for communicating intent change what users can express, understand, and repair.}

\subsection{Control Surfaces for Feed Authoring}

Existing feed controls vary substantially in how much of a user's intent they make explicit. Lightweight feedback such as ``show more,'' ``show less,'' or ``not interested'' is easy to provide while browsing, but it generally treats a post as a single object even when the user's reaction concerns its topic, author, source, tone, or another property. \citet{fengMappingDesignSpace2024}, for example, found that people evaluate feed content using varied and nuanced signals that current feedback mechanisms do not always capture.

Structured controls expose more of the reason behind a preference. Bluesky supports algorithmic choice through third-party feeds~\cite{theblueskyteamAlgorithmicChoiceCustom2023}, while tools such as SkyFeed expose filtering, regular expressions, and source composition~\cite{FeaturedCommunityProject2023}. Research systems have likewise explored rules and example labeling for constructing personal content classifiers~\cite{wangEndUserAuthoring2025}. Greater expressiveness, however, can introduce additional work: users may prefer less complex interfaces even when richer ones provide more control~\cite{jinHowDifferentLevels2017}, and translating an everyday preference into executable parameters can require a ``computational'' way of thinking~\cite{konigChallengesEnablingUser2022}. Work on interactive recommenders also suggests that these interaction forms need not be mutually exclusive. Hybrid critiquing systems, for instance, combine system-proposed options with user-initiated critiques, allowing people to move between low-effort suggestions and more self-directed specification~\cite{10.1145/1216295.1216308}.

Large language models introduce another control surface by letting people state intent in ordinary language. \textsc{Bonsai} translates natural-language goals into a transparent process for sourcing, curating, and ranking a Bluesky feed~\cite{10.1145/3772318.3791855}, while other recent systems use language models to elicit preferences users may not initially articulate or to maintain semantic profiles that support conversational feed steering~\cite{10.1145/3772318.3791569,xu2026shape}. These approaches reduce the need to formulate formal rules, but they do not eliminate the work of specifying and validating intent. Bonsai participants, for example, valued transparency while also finding intentional feed construction more effortful than familiar feed use. Prior work therefore presents post-level feedback, structured rules, and natural language as different ways of authoring personalization, and direct comparisons suggest that each can have contextual strengths~\cite{wangEndUserAuthoring2025}.

\textit{Our work brings these mechanisms together within a shared feed-authoring system and studies them as nested control surfaces. This allows us to ask not only which mechanisms people prefer, but what they use each one for, what additional expressiveness costs, and how they move among surfaces while browsing and curating.}

\subsection{Teaching, Inspection, and Repair}

Interactive machine teaching (IMT) frames users as teachers who iteratively communicate knowledge to and correct a learned system~\cite{simardMachineTeachingNew2017,ramosInteractiveMachineTeaching2020}. Applied to social media, this perspective emphasizes that teaching occurs alongside ordinary browsing rather than as a one-time configuration task. \citet{fengMappingDesignSpace2024} argue that feed controls should remain available without becoming intrusive and should support interaction across short and long timescales. Similarly, COMPASS supports continuous feed alignment through lightweight in-situ reflections during everyday browsing, helping users articulate evolving preferences without requiring them to leave the feed for a separate configuration task~\cite{barua2026compass}. Discoverability is also important: users may not seek out controls whose existence or possible outcomes they do not know about~\cite{hsuAwarenessNavigationUse2020}.

Expression alone, however, does not guarantee control. Users also need to understand what the system has recorded and correct it when its
interpretation diverges from their intent. This distinction also appears in broader human--AI interaction guidance. \citet{10.1145/3290605.3300233} identify efficient correction, explanations of why a system acted, granular feedback, and conveying the consequences of user actions as separate requirements for effective interaction with adaptive systems. Work on scrutable user models has therefore explored representations that users can inspect and modify. \citet{10.1145/3331184.3331211}, for example, make the recommender's user model explicitly visible and adjustable rather than leaving users to manipulate collections of item-level feedback, while later work explores natural-language representations of interests as readable and editable user profiles~\cite{10.1145/3477495.3531873}. Related work on explanatory debugging connects explanations to actions through which users can correct system behavior, while recent work such as SemanticCommit considers how people update and resolve conflicts within persistent stores of user intent~\cite{teso2023leveraging,10.1145/3746059.3747778}.

These ideas motivate treating inspection and repair as part of the control surface rather than as diagnostics added after personalization occurs. \textit{In \system{}, post-level actions, authored rules, and model-assisted suggestions write to a shared preference store whose contents can be inspected and revised, while provenance and effect information connect stored preferences to resulting feed behavior. This architecture lets us study not only how users state preferences, but how they recognize when an interpretation has gone wrong and what they need to repair it.}

\section{\system{}: Authoring a Feed From Inside the Feed}
\label{sec:design}

\system{} is a feed authoring system built as a fork of the Bluesky web client together with a server that stores each user's feed configuration and generates feeds on request. Because Bluesky publishes an open API for fetching posts and for serving custom feeds, participants in both of our studies read real posts drawn from the live network rather than from a simulated corpus. A user can create several independent feeds, each with its own configuration.

\subsection{Design Goals}

We synthesized prior literature into three goals which guided the development of \system{} in Study 1. 

\begin{enumerate}
    \item[\textbf{(DG1)}] \textbf{Expressiveness:} Users should be able to state nuanced positive and negative preferences and produce meaningfully different feeds.
    \item[\textbf{(DG2)}] \textbf{Understandability:} Controls should use concepts people can interpret, with a clear relationship between an action and its expected effect.
    \item[\textbf{(DG3)}] \textbf{Integration:} Authoring should remain reachable from the feed that prompted it without preventing users from returning to ordinary browsing.
\end{enumerate}

\subsection{Feed Generation}

A feed is defined by three kinds of configuration. \textbf{Sources} say where posts come from, and can be the user's Following timeline, any feed published on the network, or a list. \textbf{Filters} include or exclude posts that match a predicate. \textbf{Sorting} determines the order, with a choice of interleaving posts from each source, sorting chronologically, or sorting by priority. Priority combines recency with user-set weights, so that a post's score decays with age and is raised or lowered when it matches a weighted predicate. A user who assigns a weight of $+2$ to posts mentioning HCI moves those posts up without removing anything else. \Cref{fig:feed-pipeline-comparison} (top) shows the batching pipeline used in Study 1 for feed generation, and Appendix~\ref{app:feedgen-phase1} gives the implementation details for batching and ranking.

In this first version, predicates operated on two visible features of a post, its author and the keywords it contains. This let us examine how people used controls embedded in the feed while keeping each rule directly tied to information they could see. \Cref{sec:evolution} describes the capabilities we added afterwards for the second study and why.

\subsection{Interface}

\begin{figure*}
    \centering
    \includegraphics[width=\linewidth]{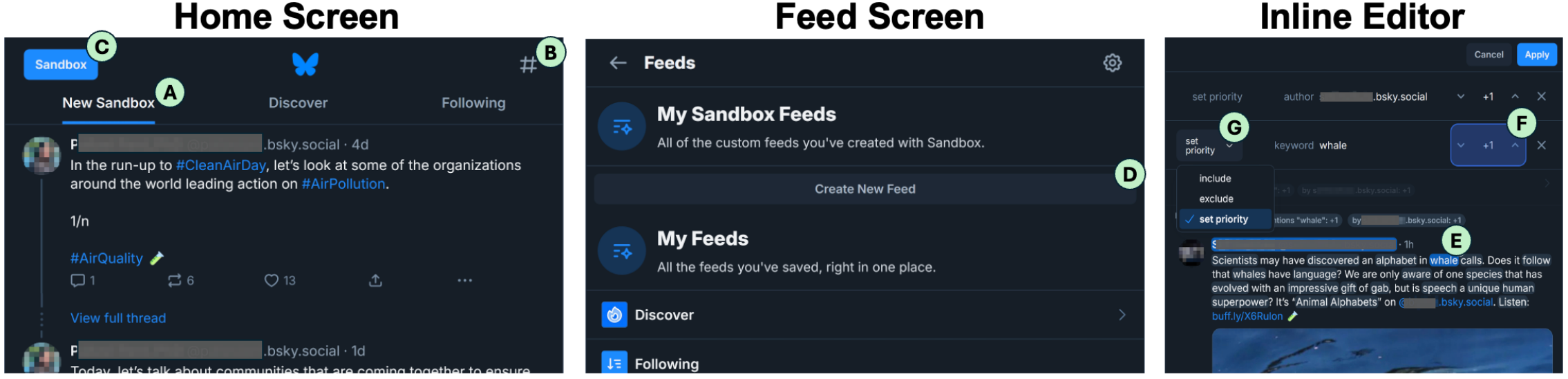}
    \caption{\textbf{Screenshots of feed navigation and inline authoring.} A custom feed appears as a tab on the Home screen (\textbf{A}). The hashtag button opens the Feeds screen (\textbf{B}), while the Sandbox button opens the controls over the current feed (\textbf{C}). Users can create a custom feed from the Feeds screen (\textbf{D}). In the inline editor, selecting an author or keyword highlights matching items in blue (\textbf{E}). Users can then adjust its priority (\textbf{F}) or turn it into an include or exclude filter (\textbf{G}).}
    \Description{Three screenshots show the Home screen, Feeds screen, and inline editor. They illustrate navigation to custom feeds and controls for selecting a visible author or keyword, adjusting its priority, or creating a filter from it.}
    \label{fig:pilot-screens}
\end{figure*}

The \textbf{feed screen} lists the custom feeds a user has created and allows them to create more. The \textbf{sandbox screen} is a dedicated authoring view where the control panel stays open and the feed refreshes as the configuration changes, which suits setting up a new feed. The \textbf{home screen} is otherwise unchanged except for a button that opens the same control panel over the feed the user is already reading, so that authoring and reading can be interleaved (see \Cref{fig:pilot-screens}).

\begin{figure*}[t]
\centering
\resizebox{\textwidth}{!}{%
\begin{tikzpicture}[
    stage/.style={
        draw,
        rounded corners=2pt,
        align=center,
        minimum height=1.05cm,
        text width=2.55cm,
        fill=black!3
    },
    io/.style={
        stage,
        rounded corners=12pt,
        fill=blue!9
    },
    check/.style={
        diamond,
        draw,
        aspect=2.1,
        align=center,
        inner sep=1pt,
        text width=1.75cm,
        fill=black!3
    },
    arrow/.style={-{Latex[length=2.2mm]}, thick},
    note/.style={font=\small, align=center},
    rowlabel/.style={font=\bfseries\large, align=right}
]

% Study 1: fetching, filtering, and sorting happen for each requested batch.
\node[rowlabel] (s1label) at (-2.2,0) {Study 1};
\node[io] (s1request) at (0.3,0) {Client requests\\feed};
\node[stage] (s1fetch) at (3.4,0) {Fetch posts\\from sources};
\node[stage] (s1filter) at (6.5,0) {Apply literal\\filters};
\node[check] (s1target) at (9.5,0) {500 posts?};
\node[stage] (s1sort) at (12.85,0) {Sort current\\batch};
\node[io] (s1return) at (16.0,0) {Send feed\\to client};

\draw[arrow] (s1request) -- (s1fetch);
\draw[arrow] (s1fetch) -- (s1filter);
\draw[arrow] (s1filter) -- (s1target);
\draw[arrow] (s1target) -- node[above,note] {yes} (s1sort);
\draw[arrow] (s1sort) -- (s1return);
\draw[arrow] (s1target.south) -- ++(0,-0.7) -| node[pos=0.28,below,note] {no} (s1fetch.south);

% Study 2: source-dependent work is separated from deterministic re-ranking.
\node[rowlabel] (s2label) at (-2.2,-3.2) {Study 2};
\node[io] (s2trigger) at (0.3,-3.2) {Sources change\\or pool expires};
\node[stage] (s2fetch) at (3.4,-3.2) {Fetch sources\\in parallel};
\node[stage] (s2pool) at (6.5,-3.2) {Interleave, deduplicate,\\and cap pool};
\node[stage,fill=green!10] (s2enrich) at (9.6,-3.2) {Add literal features,\\keywords, and\\LLM topic tags};
\node[stage,fill=green!10] (s2rank) at (12.7,-3.2) {Filter and rank\\cached pool\\deterministically};
\node[io] (s2return) at (15.8,-3.2) {Send feed and retain\\traces and effects};
\node[stage,fill=orange!12] (s2rules) at (12.7,-5.25) {User edits controls\\or preference store};

\draw[arrow] (s2trigger) -- (s2fetch);
\draw[arrow] (s2fetch) -- (s2pool);
\draw[arrow] (s2pool) -- (s2enrich);
\draw[arrow] (s2enrich) -- (s2rank);
\draw[arrow] (s2rank) -- (s2return);
\draw[arrow] (s2rules) -- (s2rank);

\draw[gray!45,thick] (-2.9,-1.65) -- (17.75,-1.65);
\end{tikzpicture}%
}
\caption{\textbf{Comparison of the feed-generation pipelines.} In Study 1, every client request fetched and filtered posts until a batch of 500 was available, then sorted that batch. Study 2 separates source-dependent pool construction from interactive ranking. It fetches sources in parallel, interleaves and deduplicates up to 2,000 posts, enriches them with literal features, keyword candidates, and language-model topic tags, and caches the resulting pool. A control change then reruns a deterministic filter-and-rank pass over that pool. The same pass produces both the served feed and the traces used for explanations and effect statistics.\vspace{-8pt}}
\Description{Two flowcharts compare feed generation. Study 1 repeatedly fetches and filters posts until reaching 500, sorts the batch, and returns it. Study 2 builds and enriches a cached candidate pool independently, then applies user controls through a deterministic ranking pass that also records explanations and effects.}
\label{fig:feed-pipeline-comparison}
\end{figure*}

\begin{figure}
    \centering
    \includegraphics[width=\linewidth]{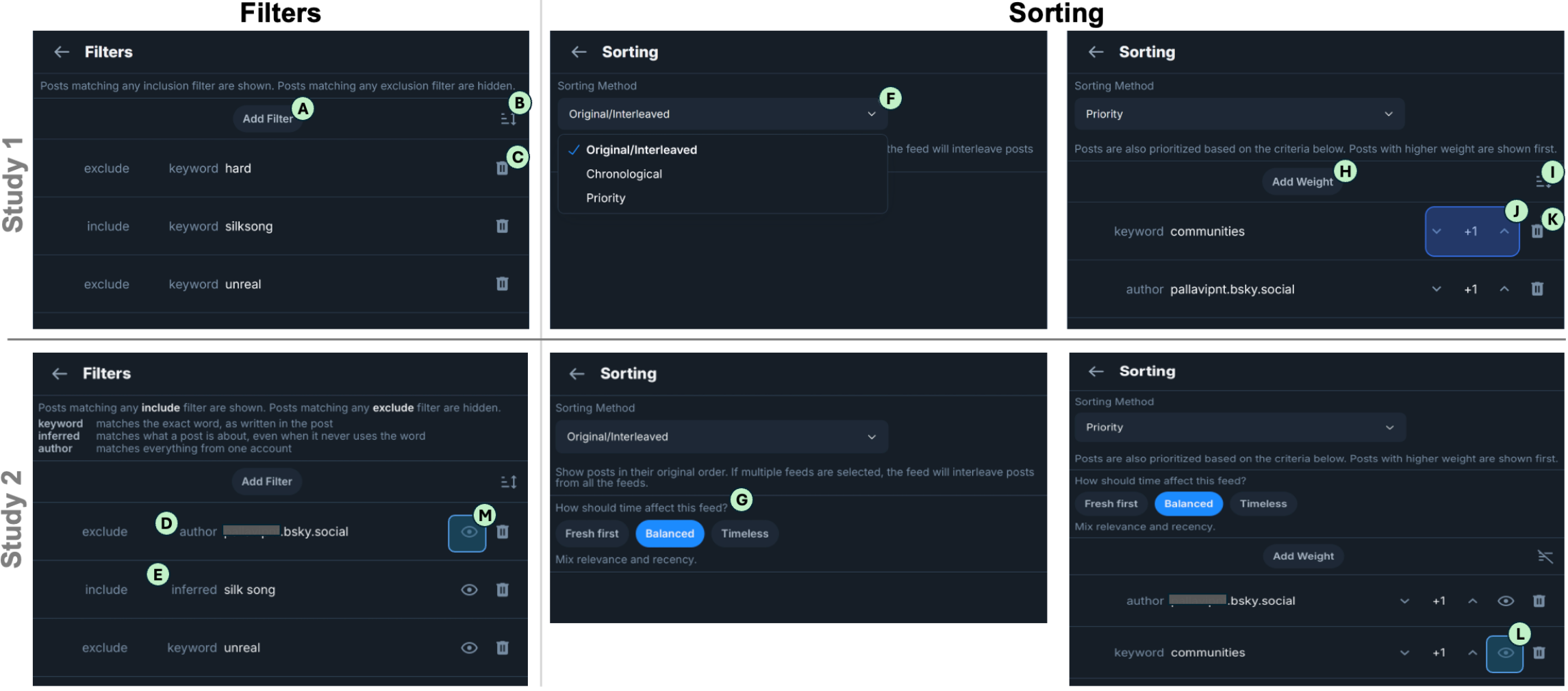}
    \caption{\textbf{Comparison of the filter and sorting editors in Study 1 (top) and the \surfaceC{} interface in Study 2 (bottom).} In the filter editor, users could add filters (\textbf{A}), sort existing filters by mode (\textbf{B}), and delete them (\textbf{C}). Study 2 distinguished literal author or keyword rules (\textbf{D}) from inferred rules that match topics by meaning (\textbf{E}). In the sorting editor, users selected a sorting method (\textbf{F}). Study 2 added control over recency (\textbf{G}). Under priority sorting, users could add weights (\textbf{H}), sort the list (\textbf{I}), adjust a weight (\textbf{J}), delete it (\textbf{K}), and under Study 2, pause a filter or sorting preference (\textbf{L}, \textbf{M}).}
    \Description{Six screenshots compare the filter and sorting editors from Study 1 with the expanded versions used in Study 2. The newer interface adds inferred topic filters, recency controls, pausing, and undo while retaining controls for adding, sorting, editing, and deleting filters and weights.}
    \label{fig:pilot_sorting_filtering}
\end{figure}

The \textbf{control panel} holds three editors. The source editor selects which feeds and lists to draw from, and supports search. The filter editor (\Cref{fig:pilot_sorting_filtering}) shows one row per predicate, each with a mode, a feature type, and a value, all of which are editable in place. The sorting editor (\Cref{fig:pilot_sorting_filtering}) selects the sorting method and, under priority sorting, shows a table of weighted predicates. Both tables can be sorted, which gives the user an overview of what their feed is currently doing. While priority sorting is active, each post carries a breakdown of how its score was computed.

The \textbf{inline editor} (\Cref{fig:pilot-screens} right) is the part of the design that most directly follows the integration design goal. When the control panel is open, the author of each post and salient keywords within it become clickable. Selecting one opens a small editor prefilled with that author or keyword, from which the user can raise or lower its priority or turn it into a filter. The intent is that a reaction to a specific post can become a rule without the user having to recall and retype the term that provoked it.

\section{Study 1: Exploring Direct Feed Authoring}
\label{sec:study1-method}

Our first study asked what people attempt to express when direct controls are placed in front of them (RQ1a), where those attempts exceed what the controls can carry (RQ1b), and how authoring sits alongside ordinary reading (RQ1c). The whole session was designed for exploration. Participants chose their own topic and their own sources and used whichever controls they liked in their preferred order.

\subsection{Recruitment, Study Design, and Analysis}

We recruited seven participants by posting in several university Slack workspaces. Five were female and two male, and three had used Bluesky before. We refer to them as F1 through F7, to keep them distinct from the participants in our second study. The study was approved by our institution's Institutional Review Board (IRB). Sessions lasted about an hour and participants received a \$20 gift code. 

The interviewer first walked each participant through creating a feed, adding sources, using the inline editor, and opening the filter and sorting editors from both the sandbox and home screens. The controls were then reset, and participants built a feed on a topic they cared about while thinking aloud. After 15 to 20 minutes they were asked to try any features they had not yet used and to build a second feed, and the session closed with a semi-structured interview. The first author then analyzed observations and interview responses using inductive thematic analysis~\cite{braun2006using}. Since participants chose their own topics and sources, the feeds differed substantially in content, so we report what people tried to do with the controls rather than comparing outcomes across participants. The results are the design input for the system described in \Cref{sec:evolution}.

\subsection{RQ1a: What People Expressed, and Which Controls They Reached For}

The available controls supported various workflows, and participants divided them along a line the design had not drawn for them. Filters were treated as statements of certainty and sorting weights as statements of degree. F6 preferred filters because their effect is \textit{``binary,''} and used inclusion filters almost as a search tool. F2 avoided filters for the opposite reason, wanting to reduce the risk of missing something, and worked almost entirely through weights because they were not \textit{``all or nothing.'}' The remaining participants mixed the two. F7 summarized the range as: 

\pquote{just [filtering and sorting] alone, you can do a million things with them.}{F7}

The two filter modes were used at different moments. Inclusion filters were typically authored deliberately and then revised through experimentation, as when F4 set a filter on \emph{education} to narrow a science feed, then added \emph{student} to widen it again and found relevant posts they had not anticipated. Exclusion filters were almost always reactive, created immediately after an unwanted post appeared. F3 excluded \emph{AI} and \emph{GPT} on encountering generative AI content, and F6 excluded an author whose posts were opinion rather than the recommendations they were looking for. In these reactive cases participants nearly always went through the inline editor, clicking the offending keyword or author in the post itself rather than opening the filter table.

Weights were used more cautiously and with less agreement about scale. F2 adjusted only in steps of one, F3, F4, and F7 were comfortable with steps of five or more, and F6 and F7 tried extreme values to see what would happen. F1 and F7 used the per-post score breakdowns to see how their weights entered the calculation, while F3 and F6 found it hard to judge how much to adjust by, which is the first sign of a calibration problem that recurs in our second study. Participants also \textit{scoped preferences to individual feeds rather} than holding them globally, with F4 comparing this to keeping personal and professional life separate.

\begin{takeawaybox}
\textbf{\faLightbulbO ~\large Takeaway:} Filters and sorting supported genuinely different strategies, and participants divided them along a line the design had not drawn, treating filters as statements of certainty and weights as statements of degree.
\end{takeawaybox}

\subsection{RQ1b: Where Intentions Exceeded the Controls}

The clearest limit was that keywords name tokens while participants were reasoning about meanings. F4 wanted less political content, could not find keywords that captured it, and settled for lowering the priority of particular authors instead. F2 hit the same wall from the other direction and worked around it by raising the priority of what they did want. Both workarounds change the feed, but neither states the thing the participant actually meant. F7 handled the problem manually, adding \emph{sci-fi} along with its spelling variants, and F4 and F2 asked for the system to propose related terms instead.

Participants also asked for dimensions the vocabulary did not have at all. Three wanted to add individual authors as sources rather than only as filter predicates, three wanted feeds organized around images, and two wanted to filter on whether a post contained an image or a link. F6 wanted recency to matter for news but not for books and art, which is a request to control how time enters ranking rather than a request for another predicate, and F7 wanted to draw from the whole network rather than from named sources. Higher-level properties of a post, of the kind a classifier might provide, were out of reach entirely.

A second kind of limit concerned what participants could see rather than what they could say. F3, F4, and F6 all had moments where they did not know which source a post had come from, which left them unable to act on it. F6 saw a post in a language they could not read, guessed at the source feed responsible, removed it, saw no change, and moved on. The action was available, but the information needed to aim it was not.

\begin{takeawaybox}
\textbf{\faLightbulbO ~\large Takeaway:} The binding limit was not the number of controls but the gap between literal predicates and the reasons participants actually had, together with missing information about why a post was present.
\end{takeawaybox}

\subsection{RQ1c: Authoring Alongside Reading}

Participants described authoring and reading as different activities and valued being able to move between them. F2 said inline feedback felt more natural while browsing because it let them keep reading, and reserved the sandbox screen for setting up a new feed. F1 found clicking directly on an author or keyword easier than composing a rule from scratch, and F7 said that making everything clickable was what kept the process enjoyable enough to continue.

\pquote{Having everything highlight-able made for a good, easy, fun, user experience.}{F7}

The option to disengage mattered as much as the controls themselves. F3 and F6 valued being able to simply browse, and F6 noted that they often use social media to relax and would not want to think about customization in that mode, while still wanting the controls one click away. F5 took the opposite position, saying they would keep the controls open all the time. Visibility also made customization imaginable. F7 contrasted the prototype with Bluesky's existing customization options, which felt \textit{``more invisible to me, so I haven't gone through the process.''} Together, these attitudes suggest that the visibility of the controls should itself be under the user's control.

Immediate feedback sustained engagement. F2, F6, and F7 said that seeing the feed change right after an edit built their confidence and encouraged them to continue, and the effect was strongest for filters, whose consequences are obvious, and weakest for weights, whose consequences are a subtle reordering. Authoring also made participants weigh their own preferences more explicitly, as when F5 traded off a narrow focused feed against the chance of finding something new. All seven reported feeling more agency over the feed than they do on the platforms they normally use, which we take as a statement about how the interface felt during a short session rather than as evidence about outcomes over time.

\begin{takeawaybox}
\textbf{\faLightbulbO ~\large Takeaway:} Authoring and reading are different activities, and participants valued putting the controls away as much as using them. Seeing the feed change immediately after an edit is what kept them going, which argues for controls that live in the feed but can be dismissed from it.
\end{takeawaybox}

These findings established how participants used direct feed controls, where those controls did not match their intentions, and how authoring fit alongside browsing. We next explain how we expanded the interface in response and organized its capabilities into three nested control surfaces for the second study.

\section{Extending \system{} into Three Control Surfaces}
\label{sec:evolution}

Study 1 established how people used sources, filters, weights, and inline controls, where literal rules could not carry their intentions, and what information they needed to diagnose outcomes. These findings motivated a fourth design goal for the expanded system in Study 2.

\begin{enumerate}
    \item[\textbf{(DG4)}] \textbf{Inspectability and repairability:} Users should be able to inspect what the system recorded, determine how it affected the feed, and revise, pause, undo, or remove it.
\end{enumerate}

Based on this, we then expanded the interface and instantiated three nested control surfaces of \system{}. We call them \signals{}, \statements{}, and \structure{}.

The expansions followed both participant requests and changes in the design space. Participants asked for author sources, media controls, adjustable recency, and help finding related terms. Large language models (LLMs) made it practical to support the latter by interpreting intent, recommending related terms, tagging posts by inferred meaning, and translating an ordinary-language request into rules a user could review~\cite{10.1145/3772318.3791855,10.1145/3772318.3791281}. We added the following capabilities.

\para{Shared preference store:} Every new control writes to one store of sources, filters, and weights. Each entry carries state and provenance, whether it came from a post-level reaction, a manual editor, or an accepted language-model proposal. The store is the source of truth for ranking, explanation, and repair, so the surfaces can differ in what users say and see without changing what the ranker does. Appendix~\ref{app:feedgen} describes how entries are evaluated.

\para{Inferred rules and topic chips:} Literal keywords could not carry many of the intentions from Study 1. During pool construction, a language model assigns short topic tags to each post. An inferred rule first looks for an exact tag match and otherwise compares the user's term with the tags in the current pool using embedding similarity. Inclusion and weight rules use a looser bar than exclusions, because an over-broad inclusion remains visible while an over-broad exclusion silently removes content. Appendix~\ref{app:feedgen} details models, thresholds, and failure handling.

\begin{figure*}[t]
\centering
\includegraphics[width=\textwidth]{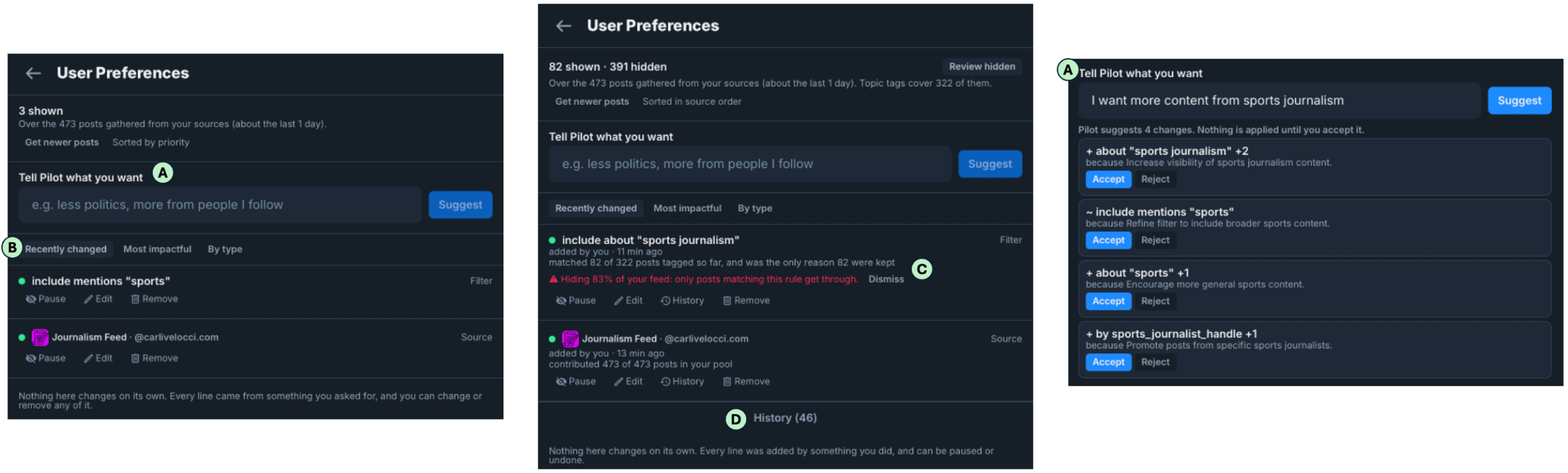}
\caption{\textbf{The \emph{User Preferences} panel in \surfaceB{} (left) and \surfaceC{} (center), together with the natural language proposal view (right).} The natural language bar (\textbf{A}) lets users describe what they want, while the preference list can be organized by recency, impact, or type (\textbf{B}). \surfaceC{} adds effect summaries and warnings about potentially problematic rules (\textbf{C}) as well as a history of earlier states (\textbf{D}). Selecting Suggest expands the bar into the view on the right, where the system proposes rules for including, avoiding, promoting, or reducing content. Users choose which proposals to accept before any are added to their preferences.}
\Description{Three screenshots show the User Preferences panel in Surface B and Surface C and the expanded natural language proposal view. The panel presents preferences in an editable list. Surface C adds effect warnings and history, while the proposal view lets users review and accept suggested rules before they are applied.}
\label{fig:user-preferences}
\end{figure*}

\para{Natural language (NL) editor:} A request bar lets users describe what they want more or less of in ordinary language. The model compiles that request into candidate filters or weights, which appear for review in the preference stores (\Cref{fig:user-preferences}). Nothing is added until the user accepts a proposal. In our study the language model helps users express intent and broaden the vocabulary available to them, but never scores or reranks posts. Appendix~\ref{app:feedgen} includes the prompts and validation steps.

\para{Machine-written-text indicator:}
Low-quality machine-written posts, often called AI slop, have become a visible part of social media feeds~\cite{yeung2026ai,10.1145/3706598.3713171}. Platforms have begun labeling such content, including Instagram's \emph{`AI Info'}\footnote{\href{https://about.fb.com/news/2024/04/metas-approach-to-labeling-ai-generated-content-and-manipulated-media/}{https://about.fb.com/news/2024/04/metas-approach-to-labeling-ai-generated-content-and-manipulated-media/}} mark and the \emph{`Made with AI'}\footnote{\href{https://piunikaweb.com/2026/03/02/x-twitter-made-with-ai-paid-partnership-labels-live/}{https://piunikaweb.com/2026/03/02/x-twitter-made-with-ai-paid-partnership-labels-live/}} disclosure on X. \system{} shows an author-level chip that estimates how machine-written an account's recent text appears. We compute it with Binoculars~\cite{hans2024spotting}. The chip is information about the content rather than a control, and therefore isn't used to filter or rank content. Appendix~\ref{app:feedgen} reports sampling details and the display thresholds.

\begin{figure*}[t]
\centering
\includegraphics[width=\textwidth]{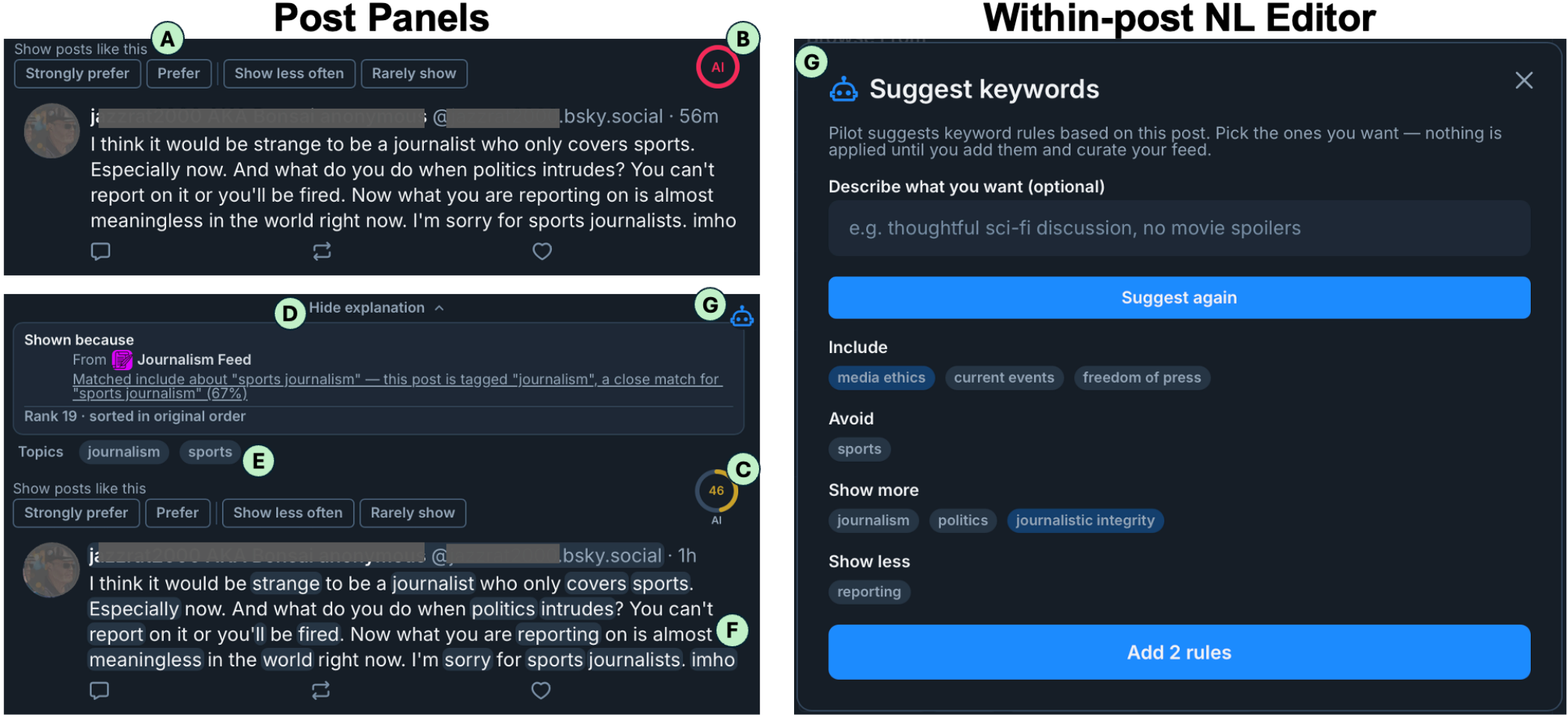}
\caption{\textbf{Comparison of controls attached to a post.} In \surfaceA{} and \surfaceB{} (top left), each post provides four preference levels (\textbf{A}) and a binary machine-written-text indicator (\textbf{B}). \surfaceC{} (bottom left) replaces this with a graded author-level chip (\textbf{C}) and adds an expandable explanation of why the post appeared (\textbf{D}), inferred topic chips (\textbf{E}), clickable keywords within the post (\textbf{F}), and a post-specific suggestion button (\textbf{G}). Opening this button displays the panel on the right, where users can describe their intent, review suggested rules, and choose which rules to add.}
\Description{Post-level interfaces from the three control surfaces. Surface A and Surface B show four preference buttons and a binary machine-written-text indicator. Surface C adds a graded chip, ranking explanation, topic and keyword controls, and a panel that proposes reviewable rules based on the post and the user's stated intent.}
\label{fig:post-level-controls}
\end{figure*}

\begin{table}[ht]
\centering\small\sffamily
\caption{\textbf{The three control surfaces of \system{} compared in Study 2.} \surfaceA{} captures directional feedback, \surfaceB{} makes stated preferences visible and editable, and \surfaceC{} adds finer specification, effects, explanations, and history.}\Description{Table comparing the three Study 2 control surfaces. Surface A includes source selection, per-post preference levels, and a binary machine-written-text indicator. Surface B includes all Surface A features and adds keyword and author filters, a visible preference store, editing and pausing preferences, and natural-language requests. Surface C includes all prior features and adds a graded machine-written-text indicator, inferred topic matching, numeric weights and recency controls, in-post topic and keyword chips, per-post explanations, hidden-post markers, effect statistics and warnings, and history with rollback.}
\label{tab:conditions}
\resizebox{\linewidth}{!}{
\begin{tabular}{lccc}
\rowcolor{blue!10}\textbf{Control surface} & \textbf{\signals{}} & \textbf{\statements{}} & \textbf{\structure{}} \\ \hline
Choosing sources & \checkmark & \checkmark & \checkmark \\
\rowcolor{gray!10}Per-post preference levels & \checkmark & \checkmark & \checkmark \\
Machine-written-text indicator & binary & binary & graded \\
\hline
\rowcolor{gray!10}Include and exclude filters (keyword, author) & -- & \checkmark & \checkmark \\
Visible store of stated preferences & -- & \checkmark & \checkmark \\
\rowcolor{gray!10}Editing, pausing, and removing entries & -- & \checkmark & \checkmark \\
Natural language requests & -- & \checkmark & \checkmark \\
\hline
\rowcolor{gray!10}User-selectable inferred topic matching & -- & -- & \checkmark \\
Numeric weights, sorting, and recency controls & -- & -- & \checkmark \\
\rowcolor{gray!10}Topic and keyword chips inside posts & -- & -- & \checkmark \\
Per-post explanations and hidden-post markers & -- & -- & \checkmark \\
\rowcolor{gray!10}Per-entry effect statistics and warnings & -- & -- & \checkmark \\
History and rollback & -- & -- & \checkmark \\ \hline
\end{tabular}}
\end{table}

\para{Version control:}
Users can pause, edit, or remove an individual entry, and roll the store back to an earlier state. History records how the store changed, which makes repair possible without reconstructing the session from memory. \Cref{fig:user-preferences} shows these controls in the preference panel.

To assess how different amounts of control support the design goals, we packaged these capabilities into three nested surfaces. \surfaceA{} keeps the simplest set of usable controls and no persistent inspectability. It offers source selection, per-post preference buttons, and the binary machine-written-text indicator, approximating the directional feedback available on mainstream platforms. \surfaceB{} adds a visible preference store, keyword and author filters, and the natural language bar. This combination is closer to existing social media controls plus recent language-based authoring systems such as \textsc{Bonsai}~\cite{10.1145/3772318.3791855}. \surfaceC{} includes the full set of features, adding inferred matching, weights, recency controls, in-post chips, effect summaries, explanations, hidden-post markers, and history. \Cref{fig:post-level-controls} shows how these additions appear on a post.

We also changed how the feed is generated so that these controls can be applied without refetching sources on every edit. Study 2 builds and enriches a cached candidate pool, then reruns a deterministic filter-and-rank pass over that pool. The bottom row of \Cref{fig:feed-pipeline-comparison} shows this split. All three surfaces use the same ranker from Study 1, so the comparison varies what users can express and inspect rather than the ranking algorithm. \Cref{tab:conditions} summarizes which capabilities appear on each surface. 

Although we instantiate these capabilities as three discrete surfaces for experimental comparison, we do not treat them as competing system designs or assume that the richest surface is the desired endpoint. Instead, the surfaces let us examine both what each layer contributes and whether participants allocate different kinds of work to different levels of control.

\section{Study 2: Comparing Control Surfaces}
\label{sec:study2-method}

Study 1 identified where direct feed authoring succeeded and where participants' intentions exceeded the available controls. Study 2 asks what changes when capabilities designed to address those gaps are progressively made available. We therefore put the three surfaces in front of the same participants in a single session, allowing each participant to experience and directly compare what the additional layers changed. The questions are about expression (RQ2a), cost (RQ2b), detection and repair (RQ2c), and how people allocate work across layers when all of them are available (RQ2d).

\subsection{Participants}

We recruited through Bluesky, university Slack workspaces, and Slack workspaces outside the university, using a call that described a paid study about controlling and customizing social media feeds. Eligible participants were 18 or older, fluent in English, located in the United States, had access to a computer with an internet connection, and had used a social media platform at least monthly over the past year. Interested people completed a short screening survey. Ten participants took part, eight women and two men, and each received a \$25 gift card for a session of about 75 minutes conducted over Zoom. We refer to them as P1 through P10. Two of the ten had taken part in the formative study described in \Cref{sec:study1-method}, which had used a narrower control surface and a different task, so the two studies do not draw on entirely independent samples.

Response to the Bluesky recruitment was low, and most participants came through the Slack channels. All but two work or study in a STEM field, and five participants had previously used Bluesky. This composition matters for how far the results generalize, and we return to it in \Cref{sec:limitations}. The study was approved by our institution's Institutional Review Board (IRB).

\subsection{Feed Sources}

In the formative study, participants chose their own sources and their own topic. For this study we prepared each participant's sources in advance, using their screening survey answers about what they wanted to see more and less of to select five sources for them, and every condition started from the same five.

This serves two purposes. Sessions were short and divided across three conditions, so we wanted participants to spend their time on the controls rather than on source selection. More importantly, had participants chosen different sources in each condition, differences between conditions would be confounded with differences in the underlying content. Participants remained free to change their sources. When someone asked to drop or replace one during their first condition, the interviewer allowed it and applied the same change to the remaining conditions before those began. Because sources were prepared rather than chosen, participants worked with a feed they had not assembled themselves, which is a departure from ordinary use that we accept as part of a laboratory setting.

\subsection{Design}

Each participant used all three surfaces in a within-subjects study design. \surfaceA{} came first for everyone. \surfaceB{} and \surfaceC{} were counterbalanced, so five participants worked through \surfaceA{}$\rightarrow$\surfaceB{}$\rightarrow$\surfaceC{} and five worked through \surfaceA{}$\rightarrow$\surfaceC{}$\rightarrow$\surfaceB{}. P1, P2, P5, P7, and P9 were in the first arm, and P3, P4, P6, P8, and P10 were in the second.

Our decision to hold \surfaceA{} first was deliberate. We used \surfaceA{} as an unexposed reference condition because encountering either richer surface first---with its additional ways of articulating, inspecting, and repairing preferences---could change how participants subsequently interpreted \surfaceA{}. Prior HCI experiments have similarly fixed a reference condition first when exposure to subsequent conditions could alter participants' behavior in that reference condition, while counterbalancing the remaining conditions~\cite{10.1145/1978942.1979301,10.1145/2702123.2702382}. Our screening survey further supported \surfaceA{} as the appropriate reference, since participants' prior deliberate feed customization was largely limited to keyword muting and `see more'/`see less'-style controls. We therefore counterbalanced \surfaceB{} and \surfaceC{}, while exposing participants to both an additive transition (\surfaceB{}$\rightarrow$\surfaceC{}) and a subtractive transition (\surfaceC{}$\rightarrow$\surfaceB{}). We note that this design does not eliminate order effects, and \Cref{sec:limitations} describes the carryover we observed.

\subsection{Semi-structured Interview Protocol}

Sessions began with a \textit{background interview} about participants' existing experiences with personalized feeds: what they wanted to see more or less of, how they currently responded to unwanted content, whether they had deliberately tried to curate feeds before, and their mental models of how feed recommendations were produced. Participants then completed the same task with each of the three surfaces: after briefly browsing the feed, they identified content they wanted to see more and less of and used whatever controls the surface provided to shape the feed accordingly. Participants were asked to think aloud about what they were trying to express, what they expected each control to do, whether the system had interpreted them as intended, and how they would diagnose or repair a misinterpretation. Near the end of a block, the interviewer occasionally invited participants to try a feature they had not yet used. 

\begin{figure*}[t]
\centering
\includegraphics[width=\textwidth]{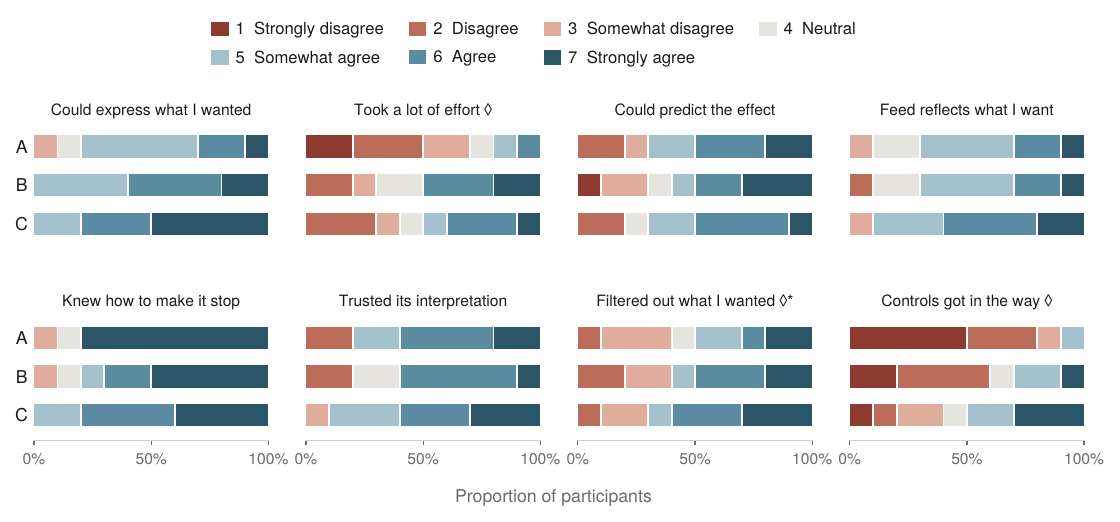}
\Description{Eight horizontal stacked bar charts, one per survey statement, each showing the distribution of seven-point agreement ratings across the three control surfaces for ten participants. Statements marked with a diamond are negatively framed. The statement about filtering out wanted content additionally carries an asterisk marking inconsistent interpretation.}
\caption{Post-surface survey responses, one panel per statement, with the three \system{} surfaces stacked within each panel. $\diamond$ marks a negatively framed statement, where agreement indicates a worse experience. $*$ marks the statement that participants interpreted in two opposite ways while thinking aloud, so its distribution should not be read as a comparison between surfaces. We report distributions rather than means because participants re-anchored the scale as the session progressed.}
\label{fig:likert-by-question}
\end{figure*}

After each surface, participants completed a short post-condition survey while thinking aloud. They rated eight statements on a seven-point scale from \textit{strongly disagree} to \textit{strongly agree}, covering whether they could express what they wanted, the effort required, predictability of changes, how well the resulting feed reflected their preferences, whether they knew how to stop unwanted content, trust in the system's interpretation, unwanted filtering, and whether the controls interfered with reading. The full item wording appears in \Cref{app:survey}. We use these ratings descriptively alongside participants' explanations rather than as inferential comparisons between surfaces. \Cref{fig:likert-by-question} and \Cref{fig:likert-by-condition} reports response distributions.

The session concluded with a \textit{exit interview} comparing the three surfaces, including which allowed participants to express the most, what effort each required, how controls fit alongside ordinary browsing, and what remained difficult to express.

\subsection{Analysis}

We analyzed the transcripts using codebook thematic analysis in a hybrid deductive and inductive form~\cite{braun2022conceptual,fereday2006demonstrating}. We started with six sensitizing domains fixed before coding and derived from the research questions and the interview protocol: \textit{(i) Expression:} what people attempt to say and what can go wrong; \textit{(ii) Transmission:} whether a stated preference gets transmitted into the ranking as intended with different controls; \textit{(iii) Verification:} whether people can tell if something worked; \textit{(iv) Effort:} what these controls cost and what makes the cost worth paying; \textit{(v) Instrument selection:} which controls and features people reach for and when; and \textit{(vi) Repair:} what people do when something goes wrong. Codes within each domain were generated inductively from an initial reading of the transcripts. Two authors discussed and finalized the codebook, and the first author then applied it across the corpus. We do not report inter-rater reliability, which is not customary for analyses of this kind~\cite{10.1145/3359174}.

\section{Findings}
\label{sec:study2-findings}

We now organize our findings by research question below.

\subsection{RQ2a: What People Expressed as the Control Surface Widened}

\subsubsection{\surfaceA{} carried judgments without their reasons:}
The per-post buttons made direction easy to state but the reason impossible to state. Eight participants produced at least one case where the basis of their judgment was not the topic of the post, and seven produced a case where the topic was right but the breadth was wrong:

\pquote{I'm just saying I prefer this post, but it doesn't know what about the post I prefer. So I'm not actually making it clear enough what I want, or I'm not able to make it clear enough what I want.}{P8}

However, the missing dimension differed from person to person. P9 marked a post down because they did not find the source credible, \textit{``so it wasn't just the topics, but also the poster.''} P8 wanted a genre, academic paper announcements, and watched the system pick up neuroscience and cognition instead. P10 objected to an outlet rather than a topic, and P6 pressed rarely show on posts written in a language they could not read, after which the system concluded they disliked art and design. P3 wanted opinion writing and diagnosed the impossibility themselves:

\pquote{No one's going to officially say the word op-ed\ldots if I say `opinion', it's gonna start looking for things with an opinion, but I just want something that's an opinion that doesn't say it's an opinion, and I don't know how to do that with the tools at hand.}{P3}

Participants then pushed non-topical intent through the topical control and absorbed the misfire, as when P10 pressed prefer on a post they did not care about because they wanted more from that publication. The scale had a gap of its own, since three participants wanted a way to say keep this as it is. P4 asked for \textit{``a neutral button\ldots I don't mind seeing this once in a while, but not all the time,''} and P10 achieved the same effect by withholding input until something recurred. 

\subsubsection{\surfaceB{} put specification in participants' hands but hit a ``literal'' ceiling:} Unlike \surfaceA{}, \surfaceB{} lets a participant write an intention down without waiting for a post to react to. Given a filter panel over keywords and authors and a request bar that turns a natural language request into candidate entries, participants generally opened the surface by stating what they wanted and then scrolled to see whether it had landed. 

However, since \surfaceB{} was based primarily on literal matching, a participant's intention survived only as far as the particular words it happens to contain. Seven participants  encountered issues similar to those seen with literal matching in Study 1, but provided additional insights. For instance, P5 asked for listings from one area of physics and saw the problem before the feed did, that the system \textit{``might be thinking that the tweet needs the word `quantum physics', which no actual paper would have.''} P8 asked for academic papers, received the keyword \emph{academic}, and objected that the system \textit{``just rephrased what I wrote,''} since they were naming a type of source rather than a word that would appear in one. P10 stated this vocabulary constraint most clearly:

\pquote{A keyword filter would either be really broad, in that the keyword is there in a lot of the content whether I want it or not, or it would be very restrictive and I would start seeing a lot fewer things\ldots I don't want it to necessarily find posts that say the word `research', I just want it to find things that are like [that].}{P10}

Neither a broader nor a narrower keyword fixes this, because what these participants wanted was a kind of post rather than a word that posts contain. \surfaceB{} also made the problem hard to diagnose. It shows the rules a participant has written but not what those rules did to the feed, so the only way to find out whether a filter was catching too much or too little was to keep scrolling and guess.

\subsubsection{\surfaceC{} changed which reasons could be named:}
The largest expression gain in \surfaceC{} came from a small feature rather than from the panels. Topic and keyword chips inside each post let participants choose which aspect of a post their preference referred to. P7 contrasted this with silent inference, since one post exposed both Elon Musk and podcasts as topics and they wanted more of one and less of the other, a distinction an automatic system would have made for them. P8 described the same shift structurally, saying the chips let them express a preference \textit{``in the content of the post, not just the whole post.''} Similarly, inferred matching addressed the literal ceiling directly, and P4 ran the contrast within a single rule. Excluding a dietary term by keyword emptied their feed, and switching the same rule to inferred matching restored it, after which they articulated the difference, that keyword exclusion \textit{``removed everything that had that,''} including posts that were useful and merely mentioned the word, when what they wanted to remove was a kind of content. They also wanted the choice per rule rather than as a global setting, since one topic warranted total exclusion and another only warranted less of it.

\subsubsection{What people still could not ask for:}

One of the key findings is that offering a control is not the same as getting it used. In \surfaceC{}, nine of ten participants still used the per-post buttons, and only four wrote a natural language request, because they had already built what they wanted with the other controls. P7 gave \surfaceA{} and \surfaceB{} the same expressiveness rating for a related reason, since the extra controls in \surfaceB{} were there but they did not expect to reach for them. 

Similarly, participants wanted to act on whether they trusted a source, on the genre of a post rather than its subject, and on the language it was written in. Six participants wanted more variety, more balance, or a steady trickle of things they had not seen before, and none of them could ask for it directly.

\pquote{It's very hard to express that hey, I want more diversity. What I can do now is just adding more keywords. But the user may not be fully aware of what they really want. So I just want the system to reflect this kind of diversity instead of letting the user explicitly say it.}{P6}

P5 wanted a part of their feed to ignore them on purpose, saying they liked \textit{``a healthy amount of maybe 10\% of the algorithm not really listening to me,''} and P10 wanted uncurated posts mixed in so that they would still run into something new. P7 mentioned having rejected a keyword feature on another platform for the same reason, that it restricts the feed so much that there is no space left for exploration. 

\begin{takeawaybox}
\textbf{\faLightbulbO ~\large Takeaway:} A wider control surface changed which reasons participants could name rather than only how much they could say, and the features that did the most work were within-post chips that let someone point at the part of a post they meant. What no surface could carry, however, was the composition of the feed as a whole.
\end{takeawaybox}

\subsection{RQ2b: What Each Layer Cost}

\begin{figure}[t]
\centering
\includegraphics[width=\linewidth]{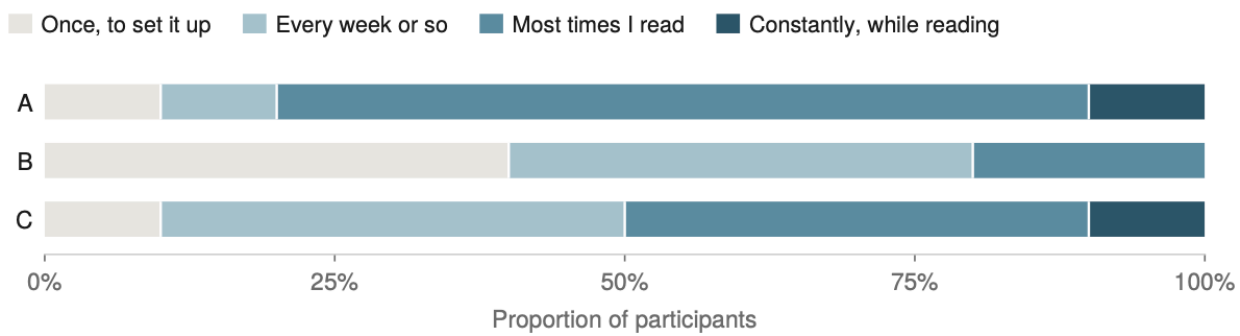}
\Description{A stacked bar chart with one bar per control surface showing how often participants said they would use the controls if this were their real feed, with categories ranging from once to set it up, through every week or so and most times I read, to constantly while reading. No participant selected never again.}
\caption{Anticipated frequency of use for each control surface. No participant chose never again. \surfaceA{} was most often imagined as something used while reading, \surfaceB{} as setup or periodic maintenance, and \surfaceC{} as a mix of the two. These are expectations stated after a short session rather than observed adoption.}
\label{fig:use-frequency}
\end{figure}

\subsubsection{Effort did not rise with capability:}
Since the surfaces are nested, we expected effort to rise with capability. However, that was not generally the case. Four participants rated a less capable surface as more demanding because cost was dominated by what happened when something went wrong. P2 rated \surfaceB{} above \surfaceC{} because \surfaceB{} \textit{``gave the appearance that it would give you the content preferences you wanted, but there were hiccups, and then you'd have to go back and adjust it manually,''} while in \surfaceC{} the chips made a change \textit{``a pretty low effort change for a pretty high impact.''} P3 also rated \surfaceB{} highest in effort, since \surfaceA{} gave them too little to think with and \surfaceC{} made control easier, while P7 rated \surfaceC{} both most expressive and least effortful because of the chips. P10 spelled out the mechanism more clearly:

\pquote{Even though I had to spend more time curating it than the previous one, I feel like it was less effort because of how directed I could be\ldots It made me think a lot more, so that is effort. It's effort that I like, it's effort that I would want.}{P10}

What participants call effort is closer to uncertainty about whether an action will do what they meant than to time spent or buttons pressed.

\subsubsection{Weights cost the most but were also missed the most:}
Numeric priority weights reliably added cost by transferring a judgment participants had little basis for making. Yet weights were also missed when removed, with P6, who met \surfaceB{} after \surfaceC{}, wanting back the weighting of each keyword. P9 and P3 gave reasons that invert one another:

\pquotetwo{The problem with this is that it leaves the burden on my calibration\ldots It totally is very, very random. So I think the burden should be on the system for that.}{P9}{I miss the weights. The biggest thing I miss is the weights\ldots something like weights makes me feel a little bit more in control, because it's pressing a plus and up and down button and not actually coming up with the word itself. Coming up with a word feels like a lot more pressure.}{P3}

Moments before their complaint, P9 had set a weight without reading the number, and P7 likewise chose a value at random. What separates the two positions is \textit{which cost each participant is trying to avoid, arithmetic or articulation}, and our study presents this as a design tension.

\subsubsection{Willingness to curate depends on the feed, not the interface:}
Five participants described curation as incompatible with why they open a feed, and P1 said they would \textit{``just close my tab and open Reddit''} if the system made social media feel like work. Four made the budget conditional on stakes, and P6 would pay a high curation cost for research papers but little for travel or food. Four also described satisficing, preferring a worse feed that costs less. A single measure of willingness to curate is therefore misleading. \Cref{fig:use-frequency} shows that \surfaceA{} was most often imagined as part of reading, \surfaceB{} as setup or weekly maintenance, and \surfaceC{} as a mix. Agreement that controls interfered with reading rose descriptively from one participant in \surfaceA{} to three in \surfaceB{} and five in \surfaceC{}.

\subsubsection{More detail moved uncertainty rather than removing it:}
More detail did not necessarily make outcomes more predictable. Seven participants agreed that they could predict a change in \surfaceA{}, six in \surfaceB{}, and seven in \surfaceC{}, but the source of uncertainty changed. In \surfaceA{} they did not know what a button meant, in \surfaceB{} they did not know what a rule would match, and in \surfaceC{} they did not know how visible entries would combine.

\pquote{Technically the user preferences area had everything listed. But it was hard to know if I made a certain change, how it would interact\ldots if I did something that caused a preference to appear, I wasn't sure how it would interact with other preferences.}{P8}

Agreement with the trust item was lower in \surfaceB{}, at six of ten, than in \surfaceA{}, at eight, or \surfaceC{}, at nine. Interviews similarly characterized \surfaceB{} as showing what it recorded but not what those records did.  

\begin{takeawaybox}
\textbf{\faLightbulbO ~\large Takeaway:} Participants treated capability and cost separately. What they priced was uncertainty about whether an action would land and the work of putting it right, so the richer surface could also be the cheaper one. The only cost that rose with every addition was interference with reading.
\end{takeawaybox}

\subsection{RQ2c: Detecting and Repairing Misalignment}

\subsubsection{Detection usually began with the feed, not the controls:}
Six participants detected a problem by reading the resulting feed, and in every case it was overshoot, where the direction was right and the magnitude was not. P6 preferred a handful of food posts and got a monoculture, and P10 observed that their feed had \textit{``very quickly captured what my stance on politics is.''} Participants then became wary of their own controls, and P10 concluded they would avoid the strongest level \textit{``considering how intensely it changes what I'm seeing,''} carrying that policy into the next condition. Five participants tested the system deliberately, flipping a preference and refreshing or issuing several signals to triangulate. This was not for want of the system speaking, since \surfaceA{} announced what it had recorded after every press, but participants split evenly on whether that notice was a window into the model or a notification to dismiss, with P9 reading it as \textit{``just a notification that this has happened, like somebody's liked our thing.''} The channel people relied on was therefore the feed itself, which arrives after the fact and shows the outcome without the cause.

\subsubsection{Legibility was wanted in aggregate and refused per post:}
Six participants rejected numeric detail at the reading surface while endorsing transparency in principle. P2 held both positions within one session.

\pquote{If I'm scrolling I'm not trying to do math\ldots [and, later, for the same `why this post?' explanation] it does give you a lot more transparency, and it helps you understand exactly how it's working.}{P2}

Five said they would open an explanation only when something looked wrong. The aggregate views, by contrast, were liked without qualification. P1, upon seeing the proportion of the candidate pool being hidden, said \textit{``this is showing me what my feed actually is,''} and P10 used the ranking of most impactful entries to establish that their filters were doing the work and their weights were not. P9 asked for exactly this before being shown it, wanting a periodic summary \textit{``not just per post, but per larger set of preferences.''} Similarly, P8, on noticing that one of their filters had no matches, said \textit{``if you don't know what it's doing, then you just start piling on filters, like what I was about to do.''} Relatedly, P3 and P8 placed their remaining uncertainty on the fact that explanations answer why a post appeared, not what would happen if a rule changed.

\subsubsection{Repair took whichever layer matched the fault, and failed repair is where trust broke:}
Once a participant knew something was wrong, the fix they chose tracked where they thought the fault was. Five edited the store directly, deleting a spurious entry, narrowing one that was too broad, or changing a weight they disagreed with, as when P10 removed an entry because \textit{``that's not really what I meant by trivia.''} Three corrected the language compiler by telling it what it had got wrong, treating it as a fallible intermediary rather than as the ranker itself. Four reset or reverted, and P8 both used the history feature and had asked for \textit{``version control of feeds''} minutes before seeing it in \surfaceC{}. Three kept repairing behaviorally, generating engagement signals rather than editing anything, which is the only option \surfaceA{} leaves open.

However, repair did not come without discomfort. Three participants saw the fault, knew the fix, and declined it, with P7 saying they would not do it if it were required every time they pressed a button. P4, having set a keyword that did not match any posts said that in the real world they would've \textit{``given up''} after their filter emptied their feed. 

Similarly, upon mistakenly clicking ``rarely show'' for the post below the intended one, leading to a post they had marked rarely show reappear, P3 said:

\pquote{If I'm genuinely telling the system rarely show, and it's still continuing to show it to me when I refresh, that would be so annoying that I would not want to come back. I would hate the experience, because it feels like my efforts are going to waste.}{P3}

They proceeded to set a tolerance of about two more occurrences before they would abandon trying altogether. The reverse also holds on the same scale, since P2's trust rose from a single accurate acknowledgment of what their press had meant. What moved trust in either direction was not how much the interface let participants say, but whether the system visibly followed through on what they had already said or the mental model they had of the system. These incidents indicate that repair is expensive, not the capability of the participants to do it.

\begin{takeawaybox}
\textbf{\faLightbulbO ~\large Takeaway:} Participants found problems by reading the feed rather than by inspecting the controls. What they needed once a problem appeared was a way to attribute the outcome to one of their own actions and undo it cheaply. Trust varied on whether the system visibly followed through, not on how much it let them say.
\end{takeawaybox}

\subsection{RQ2d: Which Layer for Which Job}

Given all three surfaces, no participant chose one. They disagreed about the individual controls while converging on the same workflow.

\subsubsection{The workflow runs coarse to fine:}
Six participants set direction with the broader controls and used per-post actions to trim, as P2 put it, treating preferences and filters as \textit{``high level, big idea sort of filtering''} and then using the per-post buttons \textit{``to fine tune''} while scrolling. P1 stated the same division as a policy about when to descend, saying the higher layers sit on top and \textit{``only if the model completely haywires or hallucinates would I manually go and do things then.''} The trimming step happened from inside the feed, since five participants preferred building rules from content already on screen, which removes the vocabulary work while keeping control over which aspect of a post becomes a rule. Four recast the per-post buttons as an exception handler for this step, and three rarely using them once richer controls existed. Filters occupied the far end, with three participants reserving them for absolute removal, as when P10 would \textit{``hop onto the filters and say exclude this from my feed altogether.''} This extends the division the first study's participants drew between certainty and degree.

\subsubsection{Some participants cleaned the store before extending it:}
P8 ran a more deliberate version of the opening step, treating the preference store as something to repair before adding to.

\pquote{I started from sort of broader things, like what it was already picking up on, and refining those first\ldots clean up the preferences is the way I'd put it. Because if I just start adding preferences and the ones that are already there are already bad, then it's just going to get confused and I'm not going to get what I want.}{P8}

They applied the same discipline to escalation, exhausting the entries the system already held before touching the language bar. 

\subsubsection{Language sits at whichever end the vocabulary problem is worse:}
Usage of the natural-language based keyword suggestions is one step participants placed differently. Three treated it as the escape when keywords don't work, and P6 used filters and sorting first, resorting to language only for what neither could capture. Three put it first to avoid inventing words, and P9 found it easier \textit{``to just put those words and have AI leverage the fact that you can just put a rough sketch of what you want.''} P10 resolved the disagreement by scale rather than by preference, using structured entries \textit{``when I want to see more of a particular thing''} and language \textit{``when I'm setting up or making bigger changes across my feed.''}

\subsubsection{The user preferences would be opened occasionally, mostly to undo the quick reactions:}
Seven participants said they would use the two kinds of control at very different rates, pressing the per-post buttons constantly while reading and opening the preference store only every week or two. P8 would set theirs up once and then check on it when they started seeing things they did not want. The reason P10 gave for going back is more interesting than the frequency, since they expected their own quick reactions to make the feed worse over time. They predicted they would \textit{``[mess] up the feed myself''} through the per-post buttons and would then visit the preferences \textit{``once in a couple of days just to fix that.''} On this account the occasional visit is for cleaning up, not for adding anything new, which is the same job P8 did at the start of their session. P3 wanted the system to prompt them instead of leaving it to them to notice, asking for a periodic check that the feed had settled into a state they were happy with. \Cref{fig:use-frequency} shows the same split in how often participants expected to use each surface.

\begin{takeawaybox}
\textbf{\faLightbulbO ~\large Takeaway:} Nobody settled on a favorite surface. Users converged on a workflow that runs coarse to fine and on two timelines, a fast loop for reacting while reading and a slow loop for maintaining the preferences store.
\end{takeawaybox}

\subsection{The Machine-Written Text Indicator Shaped Decisions Without Acting on the Feed}

The machine-written text indicator appeared across all three surfaces but was informational only, i.e., it never entered filtering or ranking. Several participants had already expressed a desire to see less machine-written content during the background interview, and some subsequently used the indicator as evidence when deciding how to act on a post. P2 used \textit{rarely show} on posts carrying the binary indicator, while P4 wanted to filter directly on the score. P4 also wanted a way to contest possible misclassification. P6 similarly said that a high indicator value would make them more likely to pass over a post because they preferred writing by people.

Five participants preferred the graded indicator in \surfaceC{} to the binary version shown earlier, describing the number as less categorical:

\pquote{I do appreciate that it went from a straight red AI to a 41 AI. Like, I like the fact that it's not a blanket binary statement on this.}{P3}

Making the cue graded in \surfaceC{} reduced its appearance of certainty for some participants, but did not by itself guarantee an accurate mental model of what the value represented.

\subsection{Order Effects and What the Two Arms Show}

Because \surfaceB{} and \surfaceC{} were counterbalanced, the two arms (\surfaceA{} $\rightarrow$ \surfaceB{} $\rightarrow$ \surfaceC{} and \surfaceA{} $\rightarrow$ \surfaceC{} $\rightarrow$ \surfaceB{}) produce different evidence. Participants who met \surfaceB{} after \surfaceC{} narrated it as loss, and four of five named a removed affordance they wanted back, including inferred matching, weights, in-post highlighting, and explanation. The first arm gives more equivocal evidence about incremental gain, since three of five participants there preferred \surfaceB{}, while four of five in the second arm preferred \surfaceC{}. Since we cannot distinguish the two, we read the asymmetry in the subtractive arm as consistent with loss aversion.

\section{Discussion and Implications}

Across our two studies, usable feed control was not achieved by adding one expressive mechanism. People moved among controls that supported different parts of authoring, and the value of a control depended on what had already happened in the feed. This shifts the design problem from choosing how much control to offer toward coordinating a set of controls over time. Our findings also let us return to the design goals that guided \system{}. For \textbf{DG1 (Expressiveness)}, richer surfaces expanded the kinds of reasons participants could state, but no single mechanism captured all of their intentions. For \textbf{DG2 (Understandability)}, participants' experiences depended less on the number of controls than on whether they could predict what an action would do and see how it affected the feed. For \textbf{DG3 (Integration)}, participants distinguished between lightweight controls they could use while reading and more involved controls better suited to occasional setup or maintenance, reinforcing the need for authoring to remain accessible without dominating the browsing experience. For \textbf{DG4 (Inspectability and repairability)}, making preferences visible, editable, and reversible gave participants ways to diagnose and correct misinterpretations that lightweight feedback alone could not provide.

We now discuss challenges that our findings open for HCI and recommender-systems research.

\subsection{From Control Suites to Control Orchestration}

Participants used different controls for different kinds of preferences they wished to express. For instance, per-post controls allowed them to provide judgments but not reasons, keywords required literal predicates, and language required a model to translate intent. Each produced different levels of expression and repair. In this sense, the three surfaces functioned not only as experimental conditions but as \textit{probes of three complementary interaction modes}: lightweight reaction, explicit preference authoring, and deeper inspection and repair.

When all surfaces were available, participants set broad direction, trimmed within the feed, reserved filters for removal, and used language where structured terms were hardest to come up with. This extends \citet{wangEndUserAuthoring2025}'s finding that authoring methods have contextual strengths and also reinforces the idea of hybrid critiquing~\cite{chen2012critiquing}, while extending interactive machine teaching into a workflow for selecting among gestures~\cite{fengMappingDesignSpace2024, ramosInteractiveMachineTeaching2020}.

Complementarity of controls, however, does not mean that every control should remain visible at once. Our study showed that this might actually make the user feel that there is a transfer of work from the algorithm to them, as they now have to not just identify what they want but choose how to represent it, and anticipate how several preferences will interact. P2 described how this freedom could work against its purpose:

\pquote{It gives you a little bit too much\ldots You get hyper-specific\ldots having those preferences might actually be counterproductive.}{P2}

This tension creates the challenge of \emph{control orchestration}, which concerns which control should appear, when it should appear, and how much initiative the system should take. A system built on the principle of mixed-initiatives~\cite{10.1145/302979.303030} might offer an aspect selector for an ambiguous reaction, inferred matching after a literal miss, or the preference store after quick reactions accumulate. Connecting to recent literature, the idea of general user models~\cite{10.1145/3746059.3747722} or behavior latticing~\cite{zhao2026behavior} could be extended to developing user models of users’ feed-curation behavior that, over time, learn to infer user motivations around using different controls to express different things. However, our study provides preliminary evidence that orchestration could recreate the intent failures it aims to solve, so suggestions should remain dismissible and explain why they appeared.

Future research could therefore study whether such a system-initiated intervention matched the user's goal, whether they accepted or rejected it, and whether the resulting preference remained consistent across different workarounds. From an interface point of view, participants largely appreciating the \textit{User Preferences} panel suggests that such systems could write to one inspectable representation, while showing provenance and allowing removal, to help turn a collection of features into a coherent feed authoring environment.

\subsection{Feed Authoring Creates Preference Debt}

Our study showed that direct controls make the act of curation cheap but can create work later. Many episodes of repair highlighted that reactions may produce broader or narrower preferences than intended, literal rules may never match, and reasonable entries may interact unexpectedly. For example, one participant expected quick reactions to \textit{``mess up''} the feed and require periodic repair, while another cleaned the store before adding anything. We use \emph{preference debt} to describe the maintenance burden that accumulates when stored signals and rules no longer represent what a user wants once their combined effects become visible.

Prior HCI research has shown that scrutable user models let people inspect and correct personalization profiles~\cite{10.1145/2395123.2395129}, but our findings shift attention from inspectability and visibility to maintenance. Systems must help users find stale, conflicting, or ineffective entries without reconstructing the feed's history. Our history rollback feature, which one participant used, acts as a prototype for this, while other recent work such as SemanticCommit demonstrates this kind of conflict detection and impact analysis for AI memory~\cite{10.1145/3746059.3747778}.

Preference stores therefore need lifecycle mechanisms. Entries can carry source, scope, age, and observed effect, while systems flag contradictions, merge repeated signals, expire temporary preferences, and confirm old ones. Moreover, feed-specific scope would matter for these mechanisms because participants separated personal and professional feeds and accepted different curation costs for work and leisure. Finally, maintenance must also fit around reading. Five participants proposed a bounded curation mode, while agreement that controls interfered with reading rose from one participant in \surfaceA{} to three in \surfaceB{} and five in \surfaceC{}. This sharpens the principle that teaching should remain available but unobtrusive~\cite{fengMappingDesignSpace2024} and could take the form of lightweight reflection during browsing~\cite{barua2026compass}. 

\subsection{Toward Agentic Curation With an Inspectable Memory}

Participants in our study also wanted to act on fuzzy properties such as credibility, delegate tasks instead of specifying rules, and move numeric calibration onto the system. Many of them also wanted diversity, balance, or calibrated `novelty.' However, all our usable controls acted on individual kinds of content rather than feed composition as a whole. These requests are difficult for rule editors but plausible for generative agents that interpret evolving goals and act repeatedly on a user's behalf~\cite{zhao2026behavior}.

Recent systems already combine persistent semantic profiles with agentic reranking~\cite{xu2026shape}, elicit overlooked feed preferences~\cite{10.1145/3772318.3791569}, and plan and curate feeds from stated intent~\cite{10.1145/3772318.3791855}. Models can also learn personalized representations of individual or community assessments of information accuracy~\cite{jahanbakhshExploringUsePersonalized2023,goyal2026vastu}. The open problem is therefore not only whether agents can curate, but how people govern what they remember and do.

This is a challenging problem as showing users the raw model state will not be enough. Agents may rely on latent representations, but users need editable commitments stating what the agent believes it should do, with provenance, scope, age, and observed effects. Users should be able to pause, revise, delete, and revert them, while aggregate summaries precede per-post detail. RECALLbot demonstrates editable memory cards~\cite{10.1145/3772318.3790714}. Feed curation additionally requires these visible memories to govern ranking faithfully rather than merely summarize it. This could create a new evaluation target of \emph{memory edit fidelity}, which asks whether changing a visible commitment reliably changes later rankings and explanations.

Future studies can hold agent and ranker quality fixed while comparing opaque, readable, and editable memories, measuring whether edits change later behavior and whether users can repair intentionally planted errors. Interfaces can also expose distributional goals such as reserving unfamiliar content, balancing viewpoints, or limiting preference dominance. Exploration and bridging already appear as recommendation objectives \cite{mcinerneyExploreExploitExplain2018, ovadyaBridgingSystemsOpen2023}, and HCI researchers should work to make these policies understandable and testable from the feed.

Finally, an agent's memory can also concentrate sensitive information about the user over time. Prior work has shown that natural language profiles can help make interests portable~\cite{10.1145/3477495.3531873}, and that decentralized systems such as Bluesky aim to provide portable accounts and independent feeds~\cite{10.1145/3694809.3700740}. This combination however could raise questions about memory ownership. Platforms may also benefit from keeping preference models proprietary, making portability and deletion questions of governance rather than interface design alone.

\subsection{Limitations and Future Work}
\label{sec:limitations}

Since exposure was brief, we cannot make claims about long-term calibration, preference drift, habituation, or adoption. Most participants completed two curation loops per condition and some completed several. A longitudinal deployment is needed to test how the workflows participants described evolve with increased familiarity, accumulated preferences, and repeated maintenance over time. Moreover, the comparison is order-sensitive because \surfaceA{} came first for everyone, while only \surfaceB{} and \surfaceC{} were counterbalanced. Participants transferred strategies across conditions and sometimes re-anchored earlier ratings. Although a between-subjects study would reduce carryover, it would forgo the direct within-participant comparisons that were central to understanding how people experienced capabilities being added or removed. Furthermore, both studies had small samples, two participants took part in both studies, and all but three participants across the two studies worked or studied in STEM. Participants in Study 2 also worked from prepared sources rather than feeds they had assembled themselves, which reduces ecological validity even though it allowed us to hold the underlying content sources approximately constant across conditions. Finally, we studied what people could express, verify, and repair rather than the objective quality of the resulting feeds. The available content pool could therefore prevent a desired outcome even when a control behaved as intended. Future deployments should evaluate both the evolution of these interaction practices over time and whether user-authored preferences produce sustained changes in feed quality.

\section{Conclusion}
In this work, we built \system{}, a working feed-authoring system on Bluesky, to study how people can more directly express, inspect, and revise the preferences that shape their social media feeds. Across two studies, we found that usable feed control is not simply a matter of providing more expressive mechanisms. Different controls supported different kinds of work: lightweight reactions helped participants steer while browsing, explicit rules and natural language supported deliberate specification, and richer inspection mechanisms helped them understand and repair the preferences accumulated over time. Greater capability also did not uniformly produce greater effort, as participants' experiences depended on whether they could understand what the system had recorded and verify its effects. Rather than converging on a single preferred level of control, participants moved among these mechanisms across tasks and timescales. Our findings therefore shift the design question from \textit{how much control should users receive?} toward \textit{how should different forms of control be orchestrated over time?} We argue that future feed systems should treat preference expression, inspection, and repair as connected parts of an ongoing interaction between people and the systems that curate what they see.

\bibliographystyle{ACM-Reference-Format}
\bibliography{references}

\appendix
%TC:ignore
\section{System Implementation Details}
\label{app:feedgen}

\subsection{Study 1 Batch Pipeline}\label{app:feedgen-phase1}

In Study 1, each client request initiated feed construction. The server fetched up to 100 posts at a time from each configured Bluesky source, applied the current filters, and repeated this process until it accumulated a batch of 500 posts. Inclusion filters were combined with OR, so a post remained if it matched any inclusion rule, while a match on any exclusion rule removed it. The surviving batch was then interleaved, ordered chronologically, or sorted by priority and returned to the client.

Under priority sorting, a post's score was
\[
    \operatorname{Priority}(p) =
    -g \ln\!\left(2 + \operatorname{Age}(p)\right) + W(p),
\]
where $\operatorname{Age}(p)$ is the post's age in hours, $g$ controls the effect of recency, and $W(p)$ is the sum of weights for predicates the post matches. With no matching weights, the order is determined by recency. Because Bluesky sources can only be traversed sequentially using cursors, sorting was correct within a batch but not necessarily across batches. A batch of 500 delayed this boundary while keeping generation under roughly ten seconds in our deployment.

\subsection{Study 2 Candidate Pool and Ranking}\label{app:feedgen-phase2-pool}

Study 2 separated slow, source-dependent processing from interactive ranking, as summarized in \Cref{fig:feed-pipeline-comparison}. Each feed used at most five sources. The server fetched sources in parallel, interleaved them so that one busy source could not occupy the pool, removed duplicate post URIs, and retained up to 2,000 posts with no more than 600 from one source. Posts explicitly tagged only in languages other than English were omitted, while untagged and multilingual posts that included English were retained. The pool was cached for two hours and could be topped up as new posts arrived.

Before caching, the server extracted literal keyword candidates and media properties and added inferred topic tags. A background process also computed the author-level machine-written-text indicator. These annotations belong to the candidate pool and are not recomputed when a participant changes a rule.

Ranking is a pure function of the cached pool, active preference entries, sorting configuration, and evaluation time. Exclusion filters are applied first. Inclusion filters are then combined with OR, and surviving posts are ordered by the selected interleaved, chronological, or priority method. The priority score retains the Study 1 form above. Study 2 exposes three values of $g$: 2.8 for \emph{Fresh first}, 1.8 for \emph{Balanced}, and 0.5 for \emph{Timeless}. Given the same pool, entries, configuration, and time, the result is deterministic.

The ranking pass emits a trace for every candidate rather than reconstructing an explanation afterwards. Each trace records the source, matched inclusion and exclusion rules, weight contributions, age term, final score, and whether the post was shown or hidden. The served feed, per-post explanations, hidden-post markers, and per-entry effect statistics are all derived from these traces. Consequently, the explanation and the displayed outcome cannot be produced by different decision paths.

\subsection{Literal and Semantic Matching}

Keyword predicates use case-normalized whole-word matching over the post text and available alt text. Author predicates match handles exactly, and media predicates are computed directly from the post record. These rules remain literal in every condition.

Only \emph{inferred} predicates use semantic matching. During pool construction, GPT-4o mini assigns up to five short topic tags to each post from at most 600 characters of text. Text posts are processed in batches of 15, and tags are cached by post URI for seven days. Image-only posts can be tagged from up to two low-detail images. If tagging fails, the post receives no inferred tags and feed generation continues.

The system first checks whether a user's inferred term exactly matches a tag. Otherwise, it embeds the term and the distinct tags in the current pool with \texttt{text-embedding-3-small}~\cite{openaiEmbeddingModels} using 256 dimensions. A semantic match must clear both a cosine-similarity floor and a z-score measured against that term's similarity distribution over the pool. Inclusion and weight predicates require a z-score of 2.75 and cosine similarity of at least 0.45. Exclusions use the stricter thresholds of 3.25 and 0.50 because an overly broad inclusion remains visible and correctable, while an overly broad exclusion silently removes content. Terms shorter than four characters use lexical matching only because short acronyms produced unreliable embedding similarities during calibration. Semantically matched weights are also scaled by match strength so a marginal match contributes less than an exact one.

\subsection{Machine-written-text indicator}

The machine-written-text indicator uses the model from Binoculars~\cite{hans2024spotting} as described in \Cref{sec:evolution}. Because Binoculars requires at least 64 tokens and most individual posts are shorter, the server aggregates recent text by author. It gathers up to 900 characters as a target, skips authors with fewer than 280 characters, and sends at most 4,000 characters per author to the service. Up to 400 uncached authors are processed after a pool is already available, in batches of eight.

Scoring runs asynchronously so detector latency or failure never delays the feed. Results are cached by author handle for seven days and shown on every post from that author. \surfaceA{} and \surfaceB{} display a binary marker when the transformed score exceeds 25, a deliberately permissive study threshold chosen to make the chip observable during short sessions. \surfaceC{} displays the transformed value from 0 to 100. This value is a presentation mapping of the raw Binoculars ratio, not a calibrated probability that a post was generated by a language model. A missing or insufficient sample produces no chip.

\subsection{Language Model Assistance and Prompts}\label{app:feedgen-phase2-prompts}

All language-model features used GPT-4o-mini~\cite{hellogpt4o} (\texttt{gpt-4o-mini-2024-07-18}) and returned constrained JSON. Topic tagging used temperature 0.2, natural language compilation and keyword expansion used temperature 0, and post-specific suggestions used temperature 0.3. Topic tagging runs while a candidate pool is prepared. The other calls occur only after an explicit user action. Natural language feed requests produce at most eight candidate filters or weights. Post-specific suggestions produce five to eight candidate keyword rules from a post and an optional description. Keyword expansion produces up to eight alternate spellings or closely equivalent phrases, keeping variants such as \emph{sci-fi}, \emph{scifi}, and \emph{science fiction} in one rule.

The server validates every response, drops malformed or duplicate proposals, limits model-proposed weights to the recommended range from $-3$ to $+3$, and requires the user to accept a proposal before it becomes a preference entry. Prompts prohibit inferring sensitive personal attributes. Failed calls return no suggestions rather than interrupting manual feed authoring. The language model never receives responsibility for ranking.

The boxes below preserve the instructions sent to the model while presenting the required JSON schemas in a more readable form. The topic-tagging system prompt was:

\begin{tcolorbox}[
    breakable,
    colback=black!2,
    colframe=black!35,
    title={Topic-tagging prompt},
    fonttitle=\bfseries
]
\small\ttfamily
You label social media posts with concise topic tags. For each post, return 1 to 5 short lowercase topic tags (1--3 words each) describing what the post is about, including implicit topics even if the exact words are not present. Prefer general, reusable topics over specific phrases. No hashtags, punctuation, emojis, or stopwords. Respond only with a JSON object mapping each post index to an array of tag strings. If a post has no meaningful content, map it to an empty array.
\end{tcolorbox}

For a natural language request, the user message contained the request and a compact list of existing entries. The system prompt was:

\begin{tcolorbox}[
    breakable,
    colback=black!2,
    colframe=black!35,
    title={Natural-language-to-rules prompt},
    fonttitle=\bfseries
]
\small\ttfamily
You convert a user's plain-language description of the feed they want into structured rules for a transparent feed-ranking system. You never rank or judge posts yourself. You only propose rules the user will review.

Rule kinds are filters, which include or exclude matching posts, and weights from -3 to +3. Predicates may match a keyword in post text, an inferred topic, a specific author, or a media property.

Interpret ``less X'' and ``more X'' as negative and positive weights. Interpret ``no X'' and ``hide X'' as exclusion filters, and ``only X'' as an inclusion filter. Prefer inferred topics for themes and keywords for specific words. You may propose editing or pausing an existing rule by its identifier. Never infer sensitive personal attributes. Return no more than eight proposals as JSON, each with an action, rule specification, and rationale of no more than 15 words.
\end{tcolorbox}

The post-specific keyword tool used the example post, the optional description, and existing keywords with the following prompt:

\begin{tcolorbox}[
    breakable,
    colback=black!2,
    colframe=black!35,
    title={Post-specific keyword-suggestion prompt},
    fonttitle=\bfseries
]
\small\ttfamily
Help a user curate their social media feed by proposing 5 to 8 candidate keyword rules from an optional example post and description of what they want more or less of. Each rule contains a lowercase keyword of 1--3 words, one action from include, exclude, show more, or show less, and a rationale of no more than 12 words. Include useful spelling variants, prefer reusable topical terms over rare phrases, and do not infer sensitive personal attributes. Return only JSON.
\end{tcolorbox}

Finally, the keyword-expansion prompt was:

\begin{tcolorbox}[
    breakable,
    colback=black!2,
    colframe=black!35,
    title={Keyword-expansion prompt},
    fonttitle=\bfseries
]
\small\ttfamily
Expand a search keyword into alternate spellings, common abbreviations, and closely equivalent phrases that a social media post might use for the same concept. Do not broaden the topic or return related but different subjects. Return no more than eight lowercase terms of 1--3 words without hashtags, punctuation, or emojis. Return only JSON.
\end{tcolorbox}

\section{Post-Condition Survey}
\label{app:survey}

Participants rated each of the following statements from 1, strongly disagree, to 7, strongly agree, after every condition, speaking their reasoning aloud as they answered.

\begin{enumerate}
    \item I could express what I actually wanted from my feed.
    \item Getting the feed to match what I wanted took a lot of effort. $\diamond$
    \item I could predict what would happen when I made a change.
    \item The feed I ended up with reflects what I want to see.
    \item When something appeared that I didn't want, I knew how to make it stop.
    \item I trusted the tool to interpret what I asked for correctly.
    \item The feed filtered out things I actually wanted to see. $\diamond$*
    \item The controls got in the way when I just wanted to browse or read. $\diamond$
\end{enumerate}

A final item asked, if this were your real feed, how often would you use these controls, with the options never again, once to set it up, every week or so, most times I read, and constantly while reading.

Statements marked $\diamond$ are negatively framed, so agreement indicates a worse experience. The statement marked * was interpreted in two opposite ways by participants thinking aloud, with some answering it as written and others agreeing while describing content that was working well. We report it in the figures for completeness and exclude it from comparative claims.

\section{Survey Responses by Condition}
\label{app:likert-by-condition}

\begin{figure*}[ht]
\centering
\includegraphics[width=0.8\textwidth]{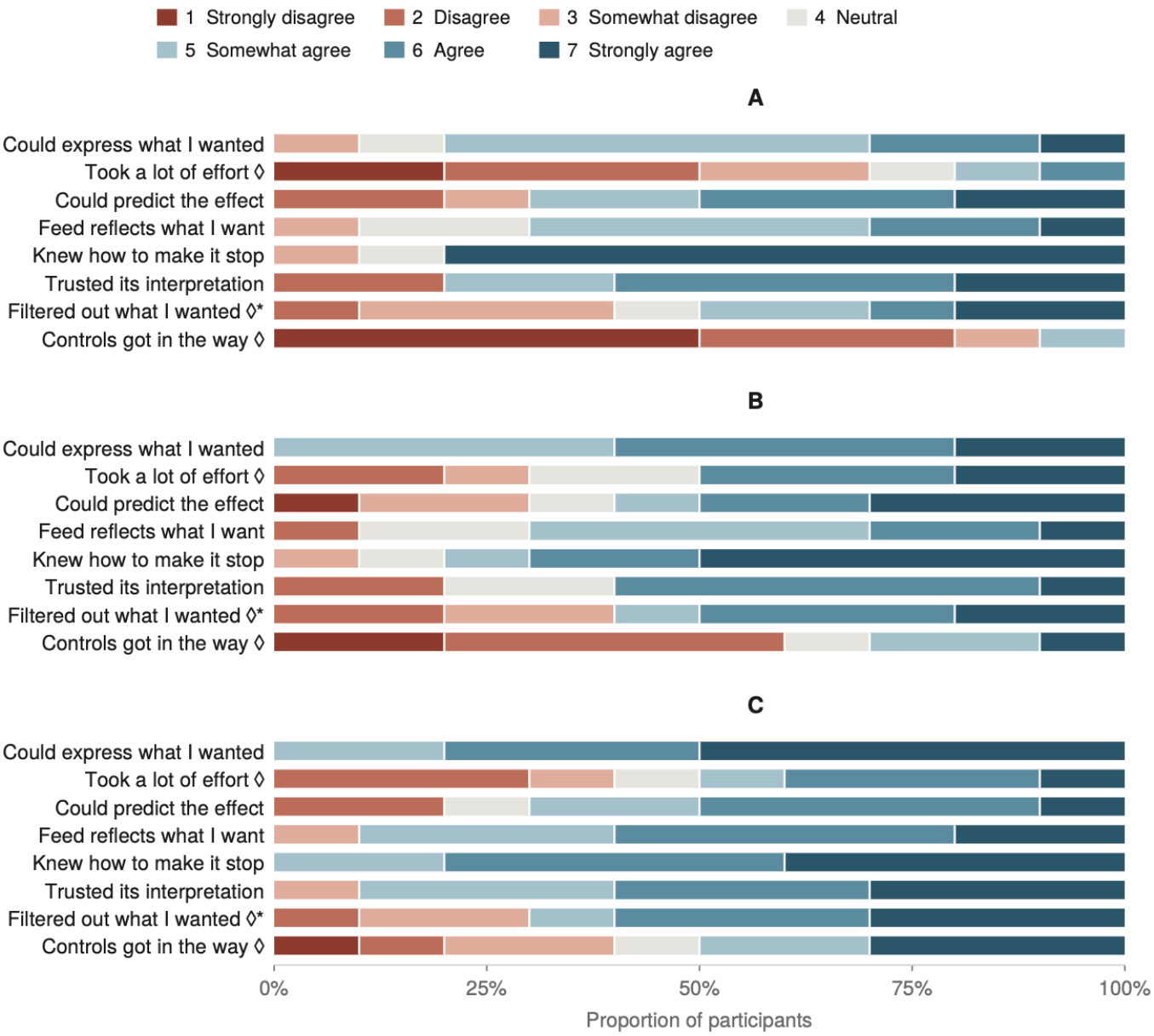}
\caption{The same survey responses grouped by condition rather than by statement. A diamond marks a negatively framed statement, where agreement indicates a worse experience, and an asterisk marks the statement that participants interpreted inconsistently.}
\Description{Three panels, one per condition, each showing stacked bar charts of seven-point agreement ratings for all eight survey statements across ten participants.}
\label{fig:likert-by-condition}
\end{figure*}
%TC:endignore

\end{document}